\documentclass[pre,aps,twocolumn,showpacs,floatfix,10pt,superscriptaddress]{revtex4-2}
\usepackage{graphicx}
\usepackage{color}
\usepackage{amsmath, amsthm, amssymb}
\usepackage{tikz}
\usepackage{tree-dvips}

\newcommand{\be}{\begin{equation}}
\newcommand{\ee}{\end{equation}}
\newcommand{\bea}{\begin{eqnarray}}
\newcommand{\eea}{\end{eqnarray}}

\begin{document}

\title{The freezing phase transition in hard core lattice gases on  triangular lattice with exclusion up to seventh next-nearest neighbor}
\author{Asweel Ahmed A Jaleel}
\email{asweel@sadakath.ac.in }
\affiliation{The Institute of Mathematical Sciences, C.I.T. Campus,
Taramani, Chennai 600113, India}
\affiliation{Homi Bhabha National Institute, Training School Complex, Anushakti Nagar, Mumbai 400094, India}
\affiliation{Department of Physics, Sadakathullah Appa College, Tirunelveli, Tamil Nadu 627011, India }
\author{Dipanjan Mandal}
\email{dipanjan.mandal@warwick.ac.uk} 
\affiliation{Department of Physics, University of Warwick, Coventry CV4 7AL, United Kingdom}
\author{Jetin E. Thomas}
\email{jetinthomas@imsc.res.in}
\affiliation{The Institute of Mathematical Sciences, C.I.T. Campus,
	Taramani, Chennai 600113, India}
\affiliation{Homi Bhabha National Institute, Training School Complex, Anushakti Nagar, Mumbai 400094, India}
\author{R. Rajesh}
\email{rrajesh@imsc.res.in}
\affiliation{The Institute of Mathematical Sciences, C.I.T. Campus,
Taramani, Chennai 600113, India}
\affiliation{Homi Bhabha National Institute, Training School Complex, Anushakti Nagar, Mumbai 400094, India}

\date{\today}

\begin{abstract}
Hard core lattice gas models are minimal models to study entropy driven phase transitions. In the  $k$-NN  lattice gas, a particle excludes all sites upto the $k$-th next-nearest neighbors from being occupied by another particle. As $k$ increases from one, it extrapolates from nearest neighbor exclusion to the hard sphere gas. In this paper, we study the model on the triangular lattice for $k\leq 7$ using a flat histogram algorithm  that includes cluster moves. Earlier studies had focused on $k\leq 3$. We show that for $4\leq k\leq 7$, the system undergoes a single phase transition from a low-density fluid phase to a high-density sublattice-ordered phase. Using partition function zeros and non-convexity properties of the entropy, we show that the transitions are discontinuous. The critical chemical potential, coexistence densities, and critical pressure are determined accurately.
\end{abstract}


\maketitle

\section{\label{sec:intro}Introduction}

Hard core lattice gases (HCLGs), a collection of  particles on lattices that interact only through excluded volume interactions, are the simplest models to show phase transitions. In particular, they are  minimal models to study fluid to solid  freezing transitions induced by increasing density~\cite{runnels1972phase,frenkel1999entropy}. Since the interaction energies are either zero or infinity, temperature plays no role in such phase transitions. The phases and the nature of the transitions depend only on the shape of the excluded volume. To understand this dependence as well as the order in which the phases occur with increasing density, the phase diagram of  many different shapes have been studied in detail in literature. Examples include  hard hexagons~\cite{1980-b-jpa-exact,baxterBook}, triangles~\cite{verberkmoes1999triangular}, squares~\cite{1967-bn-jcp-phase,2012-rd-pre-high,2016-ndr-epl-stability}, rectangles~\cite{2014-kr-pre-phase,2015-kr-epjb-phase}, tetraminos~\cite{2009-bsg-langmuir-structure}, rods~\cite{2007-gd-epl-on,2013-krds-pre-nematic}, etc.

Among the different shapes, a subclass of shapes of particular interest is the $k$-NN HCLG model in which the first $k$ next-nearest neighbors of a particle are excluded from being occupied by another particle. The case $k=1$ corresponds to nearest neighbor exclusion while the limit $k \to \infty$ corresponds to the well-studied hard sphere problem in the continuum. Also, the model can be thought of as limiting cases of spin models with long ranged interaction~\cite{1984-akg-prb-square,1985-lb-prb-phase,binder1984finite,1983-s-jpc-phase}.  Introduced in the 1950s~\cite{1958-d-nc-theoretical,1960-b-pps-lattice,1961-b-pps-lattice,1967-bn-jcp-phase,1966-bn-prl-phase,1966-bo-prl-phase,1968-OB-JCP-Traingularlattice},  the $k$-NN model, in addition  to its relevance to critical phenomena and the hard sphere gas, has found applications in diverse areas of research as well as direct experimental realizations.  Examples of applications include adsorption on surfaces~\cite{1985-twprbe-prb-two,2000-ssr-psb-phase,2001-mbr-ss-static,1998-k-jec-lattice,1985-prb-twpbe-two,2007-zbr-prb-accuracy,1989-bre-ssl-unexpected},  glass transitions~\cite{2005-eb-epl-first,2009-re-pre-critical,2003-wh-epl-glassy}, phase transitions in closely related $BM_{n}$ model~\cite{sellitto2022first}, attractive gases~\cite{orban1968jcp,prestipino2022condensation} and in combinatorial problems~\cite{1999-b-ac-planar}. Some direct experimental realizations  include porphyrins adsorbed on the Au(111) surface~\cite{gurdal2017insight,yokoyama2001nonplanar,lelaidier2017adsorption}, adsorption of chlorine on a silver (100) surface~\cite{1985-twprbe-prb-two} and adsorption of selenium on a nickel (100) surface~\cite{1985-bkuoabe-prl-phase}.

In two dimensions, the phase diagram for the  $k$-NN model for different $k$ has been studied on the square, honeycomb and triangular lattices using different techniques such as  transfer matrix calculations, high and low-density expansions, Monte Carlo simulations, etc. We summarize the known results below. 

On the square lattice, the model has been studied for $k\leq 11$, and a summary of known results may be found in Refs.~\cite{2007-fal-jcp-monte,nath2014multiple,rodrigues2021husimi}. For $k=1, 3, 6, 7, 8, 9$, the model undergoes a single phase transition from a low-density fluid phase to a high-density sublattice-ordered phase. The transition is continuous for the $1$-NN model and first order otherwise. The $2$-NN and $5$-NN models reduce to  $2\times 2$ and $3 \times 3$ hard square problems respectively and the high-density phase has columnar order. The transition belongs to the Ashkin-Teller universality class for  the $2$-NN model and is discontinuous  for the $5$-NN model. The $4$-NN, $10$-NN and $11$-NN models show multiple phase transitions. The presence of multiple transitions are due to a sliding instability present in the high-density phase and also appear for  $k>11$~\cite{nath2014multiple,2016-nr-jsm-high}.

On the honeycomb lattice, the model has been studied for $k\leq 5$~\cite{thewes2020phase,akimenko2022triangles,darjani2021glassy}. The $1$-NN model  shows a single  continuous transition, belonging to the Ising universality class, from a fluid to sublattice-ordered phase. The $2$-NN model  shows a first order transition while the $3$-NN model, surprisingly, does not undergo any transition~\cite{thewes2020phase,akimenko2022triangles}. The $4$-NN model shows a single continuous phase transition that belongs to the  three state Potts model. The $5$-NN model has been shown to undergo two discontinuous transitions. However, these transitions show non-standard scaling with system size~\cite{thewes2020phase}. 

The $k$-NN model on the triangular lattice, the main focus in this paper, has been studied for $k\leq 5$. The triangular lattice is of particular interest as it is a better approximation to the continuum than the square and honeycomb lattices. Thus, one would expect that phases seen in the continuum like the hexatic phase would be most easily seen on the triangular lattice. The $1$-NN model on the triangular lattice is the hard hexagon model. It is the only exactly solvable HCLG model that undergoes a transition from a disordered to sublattice-ordered phase which belongs to the three state Potts model~\cite{1980-b-jpa-exact,baxterBook,joyce1988hexagon}.  The $2$-NN model on triangular lattice is known to show a single second order transition which belongs to four state Potts model universality class~\cite{bartelt1984triangular,zhang2008monte,akimenko2019tensor}. The $3$-NN model has been studied in detail recently and  it undergoes a single discontinuous transition to a sublattice-ordered phase~\cite{jaleel2021hard,akimenko2019tensor,darjani2019liquid}.  Using tensor renormalization group (TRG) methods, the $4$-NN and $5$-NN models were also studied. However, the phenomenology for the $4$-NN and $5$-NN models is not well-established. The TRG results give a wide range of values for the critical chemical potential, and the nature of the transitions are also not clear~\cite{akimenko2019tensor}. Thus, compared to the square and honeycomb lattices, the $k$-NN model on the triangular lattice is less studied, and it would be of interest to determine the phase diagram for the model for $k \geq 4$.

An additional motivation for studying the $k$-NN model on the triangular lattice  comes from the algorithmic point of view. In general, Monte Carlo studies of HCLG models of particles with large excluded volume suffers from equilibration issues especially at densities close to full packing. When only  local moves are allowed, the system gets stuck in long-lived metastable states. Algorithms that include cluster moves can  overcome these issues partially. One such algorithm is  the grand canonical  transfer matrix based strip cluster update algorithm (SCUA) that updates the configurations on strips that span the lattice in one attempt~\cite{2012-krds-aipcp-monte,2013-krds-pre-nematic}. This algorithm has been very useful in obtaining the phase diagram of the $k$-NN model on square~\cite{nath2014multiple} and honeycomb~\cite{thewes2020phase} lattices as well as that of many other differently shaped particles~\cite{2013-krds-pre-nematic,2015-rdd-prl-columnar,2017-vdr-jsm-different,2018-pre-mnr-phase,vigneshwar2019phase}. However, the SCUA algorithm has difficulty in equilibrating systems which undergo a strong first order transition close to full packing~\cite{jaleel2021hard}. More recently, there have been two other algorithms that have proved useful. One is the TRG method which was used to obtain the phase diagram for the $k$-NN model on the triangular lattice for $k\leq 5$~\cite{akimenko2019tensor} and interacting hard equilateral triangles~\cite{akimenko2022triangles}. The other is the strip cluster Wang Landau (SCWL) algorithm that combines the SCUA algorithm with flat histogram methods, and thus determines the density of states for all densities~\cite{jaleel2021rejection,jaleel2021hard}. A comparison of the efficacies of the TRG method and SCWL algorithm would be useful.

In this paper, we study the $k$-NN model on the triangular lattice using flat histogram Monte Carlo simulations (SCWL algorithm) for $1\leq k \leq 7$. We benchmark our simulations against the known exact solution of the $1$-NN model. For the $2$-NN model, we obtain improved estimates for the critical density. For the $4$-NN to $7$-NN models, we show, using partition function zeros, that the system undergoes a single phase transition from a low-density fluid phase to a high-density sublattice-ordered phase. Based on non-convexity of the numerically obtained entropy, we show that the transitions are discontinuous.  The estimates for the critical parameters are summarized in Table~\ref{tab:results}.  In addition, we show that, compared to the TRG method, the SCWL algorithm is able to obtain more accurate results for larger $k$.

The remainder of the paper is organized as follows.  In Sec.~\ref{sec:model},  we define the $k$-NN model and outline the SCWL algorithm adapted to the $k$-NN model on the triangular lattice.  In Sec.~\ref{sec:results}, we first benchmark the simulations with known results from the $1$-NN and $2$-NN models. We also describe the results for the $2$-NN model as well  for $4$-NN to $7$-NN models. Section~\ref{sec:summary} contains a summary and discussion of the results.

\section{\label{sec:model} Model and Monte Carlo algorithm}

\subsection{$ k $-NN hard core lattice gas}

\begin{figure}
	\includegraphics[width=0.8\columnwidth]{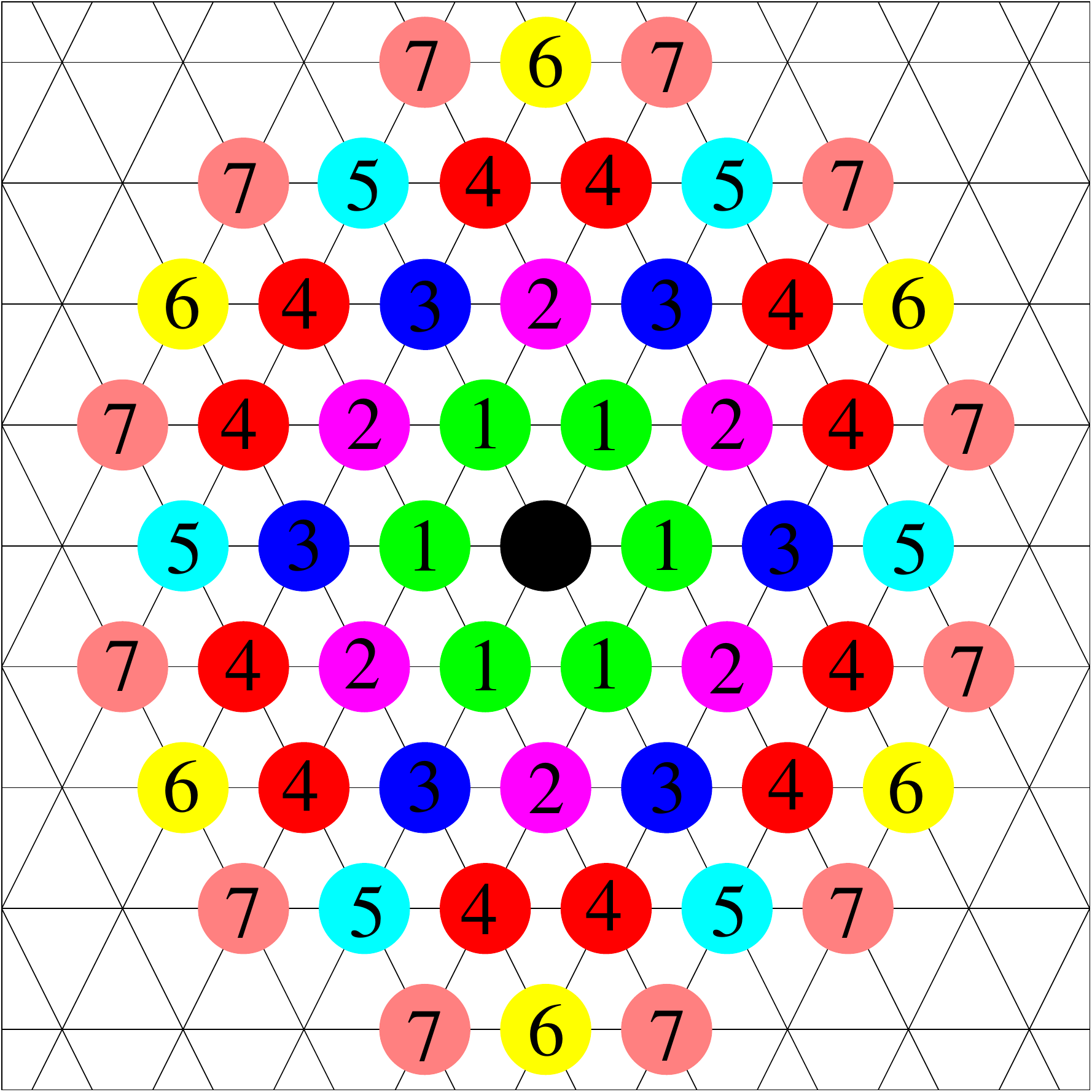}
	\caption{\label{fig:model} (Color online) The neighbors of the central site (black color) are labeled according to the distance from it. The number in the circles denote the range of exclusion $ k $. In the $k$-NN exclusion model, a particle excludes all the sites with label less than or equal to $k$ from being occupied by another particle.}
\end{figure}

Consider a $L \times L$ triangular lattice with periodic boundary conditions. A lattice site can be occupied by utmost one particle. In the $k$-NN exclusion model, a particle excludes all the sites up to the $k$-{th} nearest neighbors from being occupied by another particle. 
Figure~\ref{fig:model} shows the excluded neighbors of a particle for $k=1, 2, \ldots, 7$.  The number of excluded sites $N_{\mathrm{excl}}$ and the maximum number density $n_\mathrm{max}$ for a given $k$ are given in Table~\ref{tab:knn}. 
\begin{table}		
	\caption{For each $k$, the number of neighbors that are excluded from being occupied by a particle $ N_\mathrm{excl} $, the maximum number density $ n_\mathrm{max} $, the orientation of the rows and the range of exclusion $\delta$, as defined for the evaporation and deposition algorithm (see text in Sec.~\ref{sec:montecarlo}), are tabulated.}
	\begin{ruledtabular}
		\begin{tabular}{ccccc}
		$ 	k $& $ N_{\mathrm{excl}} $&$n_\mathrm{max}$ & orientation of rows & $ \delta$ \\
		\hline
			
			1 & 6& $ 1/3 $   &  0, $  \pi /3 $ , $ 2 \pi /3 $    &   1 			\\	
			2 &12 &  $ 1/4 $   &   0, $  \pi /3 $ , $ 2 \pi /3 $    &       1 		\\	
			3 &18&  $ 1/7 $    &    0, $  \pi /3 $ , $ 2 \pi /3 $  &       2 		\\ 	
			4 & 30& $ 1/9 $  & 0, $  \pi /3 $ , $ 2 \pi /3 $  & 2			     \\	 	
			5 & 36& $ 1/12 $  &     0, $  \pi /3 $ , $ 2 \pi /3 $  &      3  \\	
			6 & 42&  $ 1/13 $ &     $  \pi /6 $, $  \pi /2 $ , $ 5 \pi /6 $  &       2\\	 		
			7  &54& $ 1/16 $  &    0, $  \pi /3 $ , $ 2 \pi /3 $   &      3 \\	
		\end{tabular}
	\end{ruledtabular}
	\label{tab:knn}
\end{table}

\subsection{Monte Carlo Algorithm \label{sec:montecarlo}}

We study the $k$-NN model using the strip cluster Wang Landau (SCWL) algorithm introduced recently~\cite{jaleel2021hard,jaleel2021rejection}. For the sake of completeness, as well as for explaining the generalization of the algorithm to general $k$, we briefly describe the algorithm below.
 	
In SCWL, rejection free cluster moves are combined with Wang Landau flat histogram algorithm~\cite{2001-wl-prl-efficient,2001-wl-pre-determining} for estimating the density of states $g(N,L$), where $N$ is the number of particles. We  first outline the Wang Landau protocol. 

The weight of a configuration with $N$ particles is taken to be inversely proportional to $g(N,L)$. A histogram $H(N)$ records the number of times configurations with $N$ particles are visited during the simulations. Initially $H(N)=0$ and $S(N)=0$, where $S(N)$ is the entropy
\be
S(N)= \ln g(N,L).
\ee
The system evolves via an evaporation-deposition algorithm (described below). After every microstep, if there are $N$ particles, then, $ H(N)= H(N)+1$ and $ S(N)=S(N)+f $, where $ f $ is a modification factor. The system is evolved  till the histogram becomes reasonably flat, i.e., $ H_{min}(N)> c \times H_{max}(N)$, where $H_{min}$ and $H_{max}$ are  the minimum and maximum value of the histogram respectively, and $c$ is a constant. In our simulations, we choose $c=0.80$.  This is one iteration of Wang Landau algorithm. The modification factor $ f $ is halved for the next iteration and histogram $H(N)$ is reset to zero. Initially $ f $ is chosen to be equal to 1. The iterations are continued till $ f $ is less than a predefined limit, $ f_{min}$. In our simulations $ f_{min}=2^{-22} $.

We now outline the cluster move for evaporating and depositing particles~\cite{jaleel2021rejection,jaleel2021hard}. A row is selected at random. The orientations of the rows depend on $k$ and are as tabulated in Table~\ref{tab:knn}. Some of the lattice sites in the row are excluded from occupation due to particles in nearby rows. These excluded sites break up the row into open segments. We note that the orientations of the rows, as given in  Table~\ref{tab:knn}, are chosen such that the filling of these segments are independent. 

Choose a segment at random. Let this segment have $\ell$ sites. Remove all the particles in the segment and  let there be $N_0$ particles remaining in the system. Along this row, there has to be a minimum of $\delta$ empty sites between two particles. The value of $\delta$ for different $k$ are given in Table~\ref{tab:knn}. It is possible to occupy $0, 1, \ldots, n^*$ particles, where $n^*=[(\ell+\delta)/(\delta+1)]$.  The refilling is done in two steps: first we determine the number of particles $n$ that should be deposited and second we choose a random configuration from all possible ways of placing $n$ particles in $\ell$ sites. 

The probability of choosing $n$ particles to deposit is 
\begin{equation}
{\rm Prob}_o(\ell,n)=\frac{C_o(\ell,n)/g(N_0+n,L)}{\sum_{i=0}^{n^*} C_o(\ell,i)/g(N_0+i,L)},
 \label{eqn:prbo}
\end{equation}
where $C_o                                                                                                                                              (\ell,n)$ is the number of ways of placing $n$ particles on a segment of length $\ell$ sites with open boundary conditions.  If there are no sites which are excluded in an entire row, then we have to use periodic boundary condition for deposition. Then the probability ${\rm Prob}_p(\ell,n)$ of choosing $n$ particles to deposit is
\begin{equation}
{\rm Prob}_p(\ell,n)=\frac{C_p(\ell,n)/g(N_0+n,L)}{\sum_{i=0}^{n^*} C_p(\ell,i)/g(N_0+i,L)},
 \label{eqn:prbp}
\end{equation}
where $C_p(\ell,n)$ is the number of ways of placing $n$ particles on a ring of $\ell$ sites and $n^* = [\ell/(\delta+1)]$.

Determining $C_o(\ell,n)$ and $C_p(\ell,n)$ is a straightforward enumeration problem  and are given by
\bea
C_o(\ell,n)&=& \frac{(\ell+\delta-n\delta)!}{(\ell+\delta-n\delta)! n!},~n=0,1,\ldots, \left[\frac{\ell+\delta}{\delta+1} \right] \label{eqn:co}, \\
C_p(\ell,n)&=& \frac{\ell (\ell-n\delta-1)!}{(\ell-n\delta-n)! n!},~n=0,1,\ldots, \left[\frac{\ell}{\delta+1} \right].
\label{eqn:cp}
\eea

After determining $n$, we fill the segment iteratively from one end to the other. For open segments, the probability $P_o(\ell,n)$ that the first site is empty is given by
\be
P_o(\ell,n)= \frac{C_o(\ell-1,n)}{C_o(\ell,n)}=\frac{\ell+\delta- n\delta-n}{\ell+k-n\delta}.
\ee
If the first site is empty, the $\ell$ is reduced by one, keeping $n$ the same, and the step is repeated. If the first site is occupied, then $\ell \to \ell-\delta$, $n\to n-1$, and the step is repeated.
If the segment has periodic boundary conditions, the probability $P_p(\ell,n)$ that the first $ \delta $ sites are empty is given by 
\be
P_p(\ell,n)= \frac{C_o(\ell-\delta,n)}{C_p(\ell,n)}=\frac{\ell-n\delta}{\ell}.
\ee
If the first $ \delta $  sites are empty, then it reduces to an open segment of length $\ell-\delta$ with $n$ particles. Else, one of the first $ \delta $  sites, chosen at random, is occupied, and it reduces to an open segment of length $\ell-2\delta-1$ with $n-1$ particles.

$C_0(\ell,n)$, $ C_p(\ell,n) $, $ P_o(\ell,n)$ and $ P_p(\ell,n) $ do not change during the  simulation and are therefore  evaluated once in the beginning and stored as a look up table. 

The values of entropy, order parameter, susceptibility etc., reported in the paper are those calculated at the last  Wang Landau iteration.

\section{\label{sec:results}Results}

We now describe the results from the flat histogram simulations. For $k=1$, the hard hexagon model, an exact solution is available, and we benchmark our simulations against the known results. For $k=2$, we compare our results with those from earlier simulations. For $k=3$, our recent results can be found in Ref.~\cite{jaleel2021hard}. For $k=4, 5, 6, 7$, we present new results.

We start by defining different thermodynamic quantities. Knowing the density of states $g(N, L)$, the grand canonical partition function $\mathcal{L}(\mu,L)$, the pressure $P$, and the average values of any observable $\mathcal{O}$ are given by
\bea
\mathcal{L}(\mu,L) &=&\sum_{N=0}^{N_{max}} g(N,L) e^{\mu N}, \label{eqn:pf} \\
P(\mu) &=&	\frac{1}{L^2} \ln	\mathcal{L}(\mu,L), \label{eqn:pi}\\
\langle \mathcal{O} \rangle &=& \frac{\sum_{N} \mathcal{O}(N) g(N,L) e^{\mu N}}{ \mathcal{L}(\mu,L)} . \label{eqn:avgr}
\eea
The pressure in the canonical ensemble, $ \widetilde{P} (\rho)$, is calculated as~\cite{aveyard1973introduction,darjani2017extracting}
\be
\widetilde{P}(\rho)=\int_{0}^{\rho} [1-\phi(\rho)] \frac{\partial}{\partial \rho} \left[ \frac{\rho}{1-\phi(\rho)}\right] d \rho. \label{eq:blkP}
\ee
where $\phi(\rho)$ is the mean fraction of sites where a new particle cannot be placed because of exclusion due to existing particles.

For all the models, we will show that the model undergoes a single transition from a disordered fluid phase to a high-density solid phase. In the solid phase, the particles occupy one of $N_{\mathrm{sub}}$ sublattices. It is clear that $N_{\mathrm{sub}}=1/n_{\mathrm{max}}$, where the maximum number density $n_{\mathrm{max}}$ for each $k$ is as tabulated in Table~\ref{tab:knn}. We define the sublattice order parameter for a given $k$ to be 
 \begin{equation}
q_k =\left|\sum_{j=1}^{N_{\mathrm{sub}}} \rho_j\exp \left[\frac{2\pi i (j-1)}{N_{\mathrm{sub}}}\right]\right|, \label{eqn:qk}
\end{equation}
where $\rho_j$ is the density of particles in sublattice $j$. Compressibility $\kappa_k$ and susceptibility $\chi_k$ 
are 
\bea
\kappa_k&=&L^2\left[ \langle \rho_k^2 \rangle -\langle \rho_k \rangle^2 \right]\\
\chi_k &=& L^2 (\langle q_k^2 \rangle - \langle q_k \rangle^2). 
\eea
For the numerical analysis, it is useful to define an associated quantity $t_k$ defined as 
\be
t_{k}=\frac{\partial \ln \langle q_{k} \rangle}{\partial \mu}.
\label{eq:t}
\ee

From finite size scaling~\cite{fisher1983scaling,stanley1971phase,goldenfeld2018lectures} , the above thermodynamic quantities at the transition point scale with $L$ as 
\be
\begin{split}
	\kappa_{k}(\mu_c) &\sim  L^{\alpha/\nu},  \\
	\langle q_{k}(\mu_c) \rangle  &\sim L^{-\beta/\nu}, \\
	\chi_{k}(\mu_c) &\sim L^{\gamma/\nu},\\
	t_{k} &\sim L^{1/\nu},
\end{split}
\label{eq:fullscaling}
\ee 
where $\alpha$, $\beta$, $\gamma$, and $\nu$ are critical exponents. For a first order transition in two dimensions, $\nu=1/2$ and $\alpha/\nu=\beta/\nu=\gamma/\nu=2$.

Let $\mu_c$, $\rho_c$ and $P_c$ denote the critical chemical potential, critical density and critical pressure in the thermodynamic limit. For a finite size system, the deviation of these quantities from the thermodynamic limit varies as
\be
x_{c}(L)  -  x_{c} \sim  L^{-1/\nu},  ~~ x=\mu, \rho, P. \label{eq:criticalscaling}
\ee

For the first order transitions, we use the following analysis. We use the loci of the zeros of the partition function to determine the number of phase transitions. The nature of the transition is established from the angle at which the zeros approach the real axis, as well as the non-convexity properties of the entropy.   $\mu_c(L)$ is obtained from the partition function zeros, analysis of the non-convex (NC) behavior of the entropy  (see Sec.~\ref{sec:4nn} and Ref.~\cite{jaleel2021rejection,jaleel2021hard} for details), and the value of $\mu$ at which susceptibility is maximum.  The coexistence densities [$\rho_{f}(L)$ and $\rho_{s}(L)$] of first order transition are also estimated from NC analysis. 

To estimate the critical behavior accurately, the relevant thermodynamic quantities were calculated with a step size of $\Delta \mu = 10^{-5}.$ Errors in each data point are standard errors obtained from 16 independent simulations using different sequences of random numbers.  Errors in the final estimate of critical parameters are fitting errors.

\subsection{$1$-NN}

\begin{figure}
\includegraphics[width=\columnwidth]{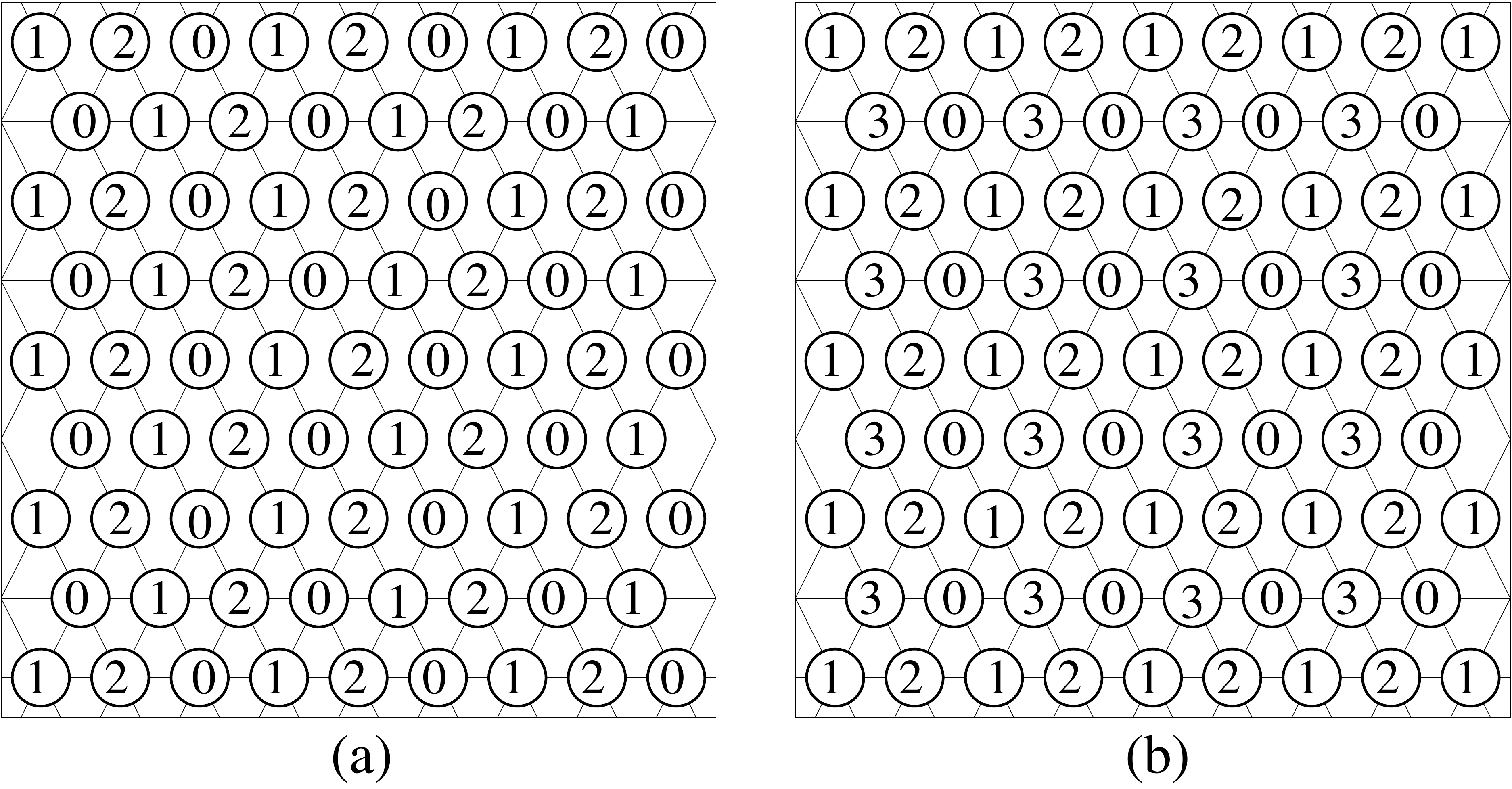}
\caption{\label{fig:sub1nn2nn}  Division of  lattice  sites into sublattices for the (a) $1$-NN model (3 sublattices) and (b) $2$-NN model (4 sublattices).  }
\end{figure}

The $1$-NN model, or the hard hexagon model, is exactly solvable~\cite{1980-b-jpa-exact,baxterBook}. There is a single continuous phase transition from a disordered phase to the sublattice phase with increasing particle density. In the ordered phase, particles preferentially occupy one of the three sublattices shown in Fig.~\ref{fig:sub1nn2nn}(a).  The critical parameters are  $\mu_c=\ln{[\frac{1}{2}(11+5\sqrt{5})]}=2.4060\dots$,  $\rho_c=\frac{3}{10}(5-\sqrt{5})=0.82917\dots$, $\alpha=1/3$, $\beta=1/9$, $\gamma=13/9$, and $\nu=5/6$~\cite{1980-b-jpa-exact}. The phase transition belongs to the three state Potts model universality class. In addition, the density and compressibility, for $\rho>\rho_c$, varies with $\mu$ as~\cite{joyce1988hexagon}
\bea
\mu(\rho)&=&\log[2(\rho-2)(3-\rho)^3]\nonumber\\
&&-\log[(27-108\rho+135\rho^2-66\rho^3+11\rho^4\nonumber)\\
&&+(-9+15\rho-5\rho^2)^{3/2}(-1+3\rho-\rho^2)^{1/2}],\label{eq:exact_mu}\\
\kappa(\rho)&=&\frac{1}{15}[(3-2\rho)(-1+3\rho-\rho^2)^{1/2}\nonumber\\
&&(-9+15\rho-5\rho^2)^{-1/2}-(-1+3\rho-\rho^2)]\label{eq:exact_compressibility},
\eea

\begin{figure}
\includegraphics[width=\columnwidth]{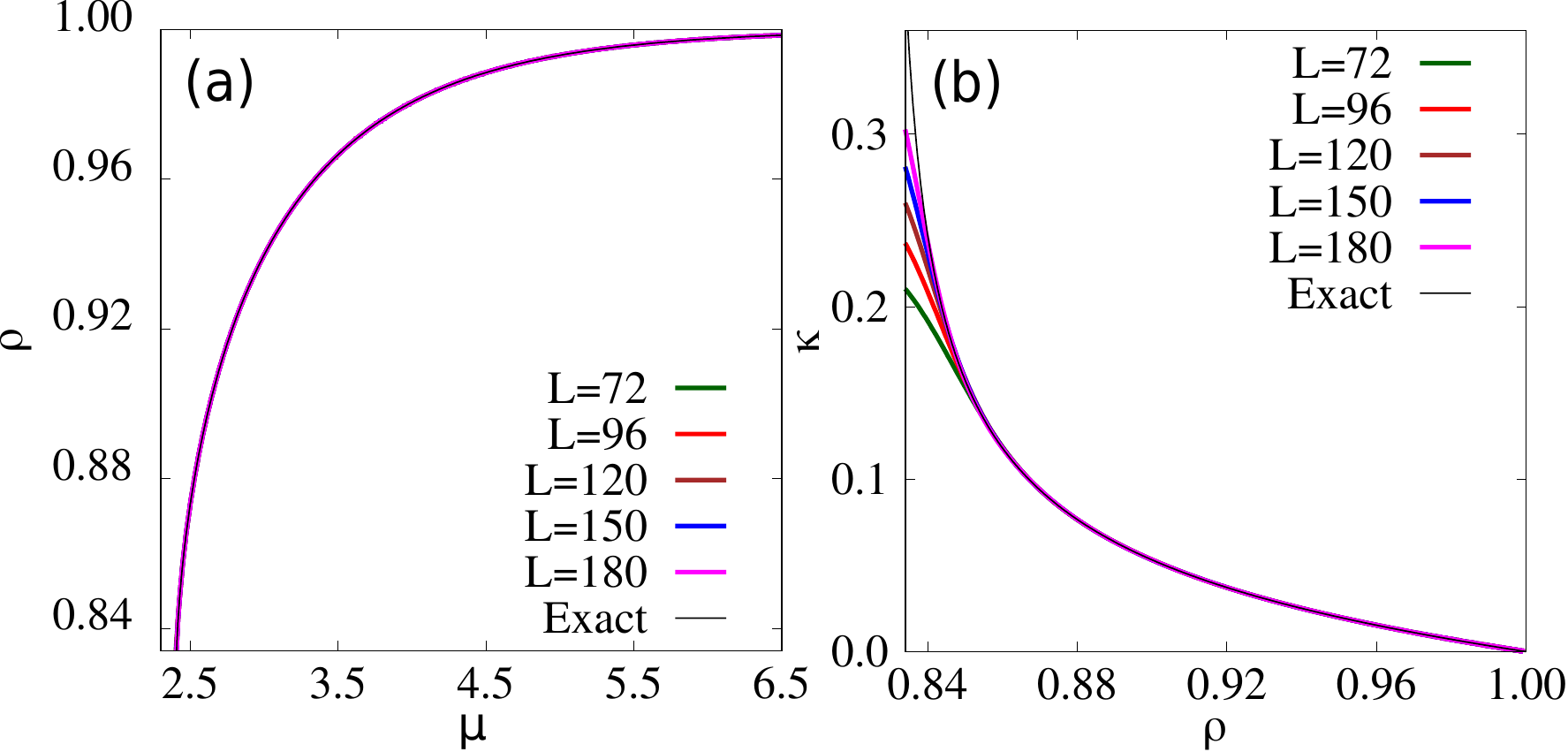}
\caption{\label{fig:joyce} (Color online)  The numerically obtained (a) density $\rho$ and (b)  compressibility $\kappa$ for the $1$-NN model are compared with the exact results as given in Eqs.~(\ref{eq:exact_mu}) and (\ref{eq:exact_compressibility}) for $\rho>\rho_c$. Data for different system sizes are shown.  
}
\end{figure}

To benchmark our flat histogram simulations, we compare the numerically obtained density and compressibility for $\rho>\rho_c$  with the exact results in Eqs.~(\ref{eq:exact_mu}) and (\ref{eq:exact_compressibility}), as shown in Fig.~\ref{fig:joyce}(a) and (b) respectively. The numerical results are in  excellent agreement with the exact results except near the critical point, where the numerical results  deviate from the exact result for the infinite system due to finite size effects. 

\begin{figure}
\includegraphics[width=\columnwidth]{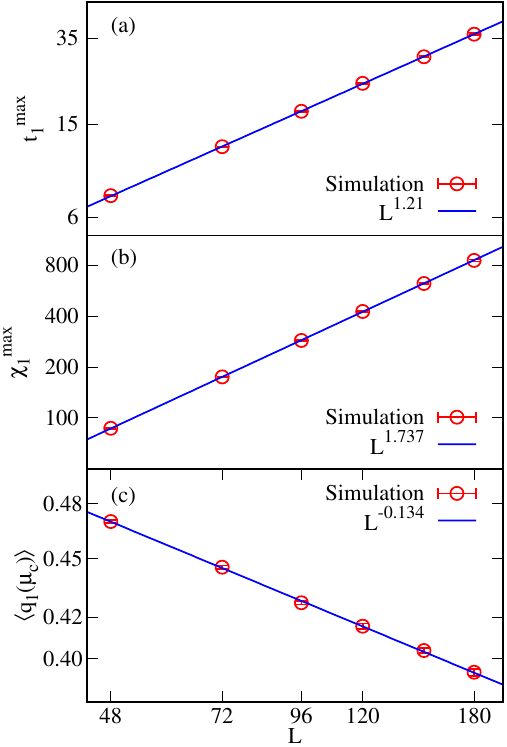}
\caption{\label{fig:1nn_exponent}(Color online) The power law  scaling of (a) $t_{1}^{\mathrm{max}}$, (b) $\chi_{1}^{\mathrm{max}}$, and (c) $\langle q_{1}  (\mu_c)\rangle$ with system size $L$ for the $1$-NN model. The solid straight lines are  best fits to the data.}
\end{figure}

\begin{figure}
\includegraphics[width=\columnwidth]{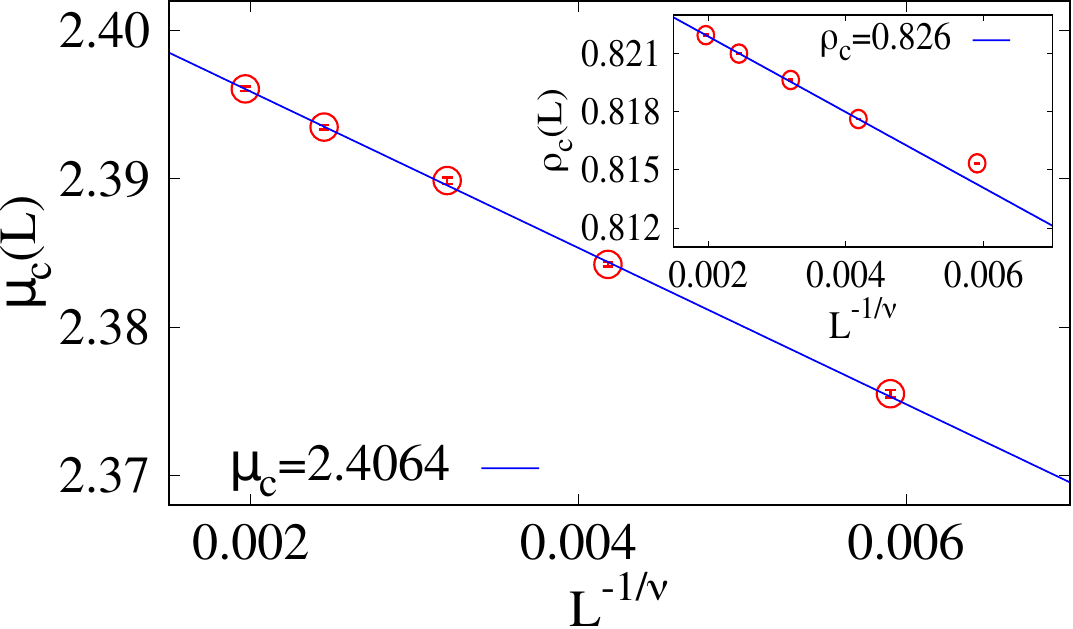}
\caption{\label{fig:muc_fit} (Color online) The extrapolation of $\mu_c(L)$ and $\rho_c(L)$ (inset) to infinite system size using the scaling in Eq.~(\ref{eq:criticalscaling}). The data are for the $1$-NN model and we have chosen  $\nu=5/6$. In the extrapolation of $\rho_c(L)$ we have used the data for only  $L\geq 96$.}
\end{figure}

We estimate the exponents using the scaling laws in Eq.~(\ref{eq:fullscaling}), as shown in Fig.~\ref{fig:1nn_exponent}. We obtain $1/\nu\approx1.21(1)$, $\beta/\nu\approx 0.134(2)$ and $\gamma/\nu\approx 1.737(6)$.  These estimates are in good agreement  with the exact value of the exponents $1/\nu=1.2$, $\beta/\nu=0.133\dots$ and $\gamma/\nu=1.733\dots$.  We determine the  critical chemical potential $\mu_{c}$ and critical density $\rho_c$ by extrapolating the $\mu_c(L)$ and $\rho_c(L)$ to infinite systems size using Eq.~(\ref{eq:criticalscaling}). $\mu_c(L)$ is identified as the $\mu$ at which $\chi$ is maximum and $\rho_c(L)$ is the corresponding density. The extrapolation is shown in  Fig.~\ref{fig:muc_fit}. We obtain $\mu_{c}\approx2.4064(6)$ and $\rho_c\approx0.826(4)$, which are in good agreement with the exact results $2.4060\dots$ and $0.82917\dots$.

In addition to these checks, we also confirm that  the entropies of the fully packed state and the state with one vacancy are reproduced correctly in the simulations. Based on the above results, we conclude that the SCWL algorithm can accurately obtain the entropy for all densities.

\subsection{\label{sec:2nn} $2$-NN}

The $2$-NN model is known to exhibit a single continuous phase transition from a disordered phase to a sublattice-ordered phase. In the ordered phase, the particles preferentially occupy one of the 4 sublattices shown in Fig.~\ref{fig:sub1nn2nn}(b). Due to the breaking of the four-fold symmetry, the phase transition is expected to  belong to universality class of the four state Potts model. The exact values of critical exponents for four state Potts model are $\alpha=2/3$, $\beta=1/12$, $\gamma=7/6$ and $\nu=2/3$~\cite{wu1982potts}.

The $2$-NN model has been studied earlier using different numerical methods. 
The known estimates for $\mu_c$ are $\approx 1.75989$ from transfer matrix scaling~\cite{bartelt1984triangular}, $1.75682(2)$ using Monte Carlo simulations~\cite{zhang2008monte}, $1.75587$ using transfer matrix methods~\cite{akimenko2019tensor} and $1.75398$ using tensor renormalization group method~\cite{akimenko2019tensor}. The corresponding values for $\rho_c$ are  $ 0.74856$~\cite{bartelt1984triangular}, and $0.72 (2)$~\cite{zhang2008monte}. The numerical estimation of the critical exponents have convergence issues due to logarithm corrections, both additive and multiplicative, that are present for the four state Potts model~\cite{salas1997logarithmic}. The earlier estimates for the critical exponents of the $2$-NN model were $\nu=0.72$ and $\beta/\nu= 1.115$ from transfer matrix scaling~\cite{bartelt1984triangular}, and $1/\nu=1.51(1)$ and $\beta/\nu=0.1257(7)$ from Monte Carlo simulations~\cite{zhang2008monte}.

\begin{figure}
\includegraphics[width=\columnwidth]{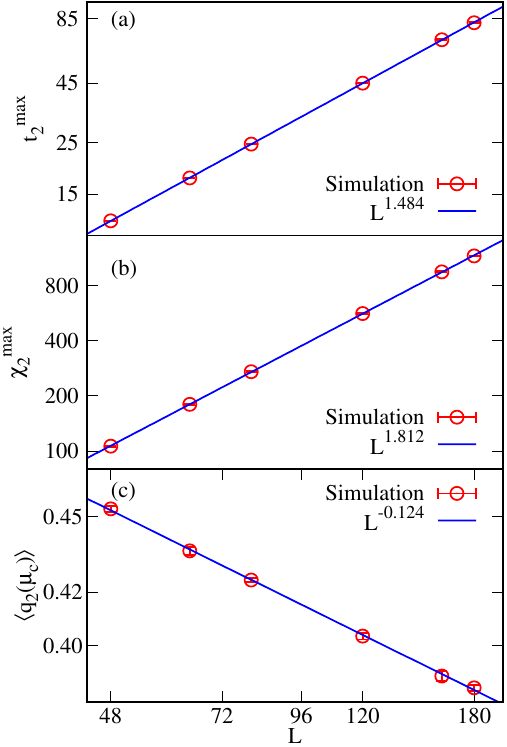}
\caption{\label{fig:2nn_exponent} The power law  scaling of (a) $t_{2}^{\mathrm{max}}$, (b) $\chi_{2}^{\mathrm{max}}$, and (c) $\langle q_{2}  (\mu_c)\rangle$ with system size $L$ for the $2$-NN model. The solid straight lines are best fits to the data.}
\end{figure}

We estimate the critical exponents for the $2$-NN model using the scaling laws in Eq.~(\ref{eq:fullscaling}), as shown in Fig.~\ref{fig:2nn_exponent}.  We obtain  $1/\nu = 1.48(2)$, $\beta/\nu\approx0.124(3)$ and $\gamma/\nu\approx1.812(8)$. The known exact value of the exponents are $1/\nu=1.5$, $\beta/\nu=0.125$ and $\gamma/\nu=1.75$. The value of $\gamma/\nu$ is overestimated, possibly because of ignoring logarithmic corrections.

\begin{figure}
\includegraphics[width=\columnwidth]{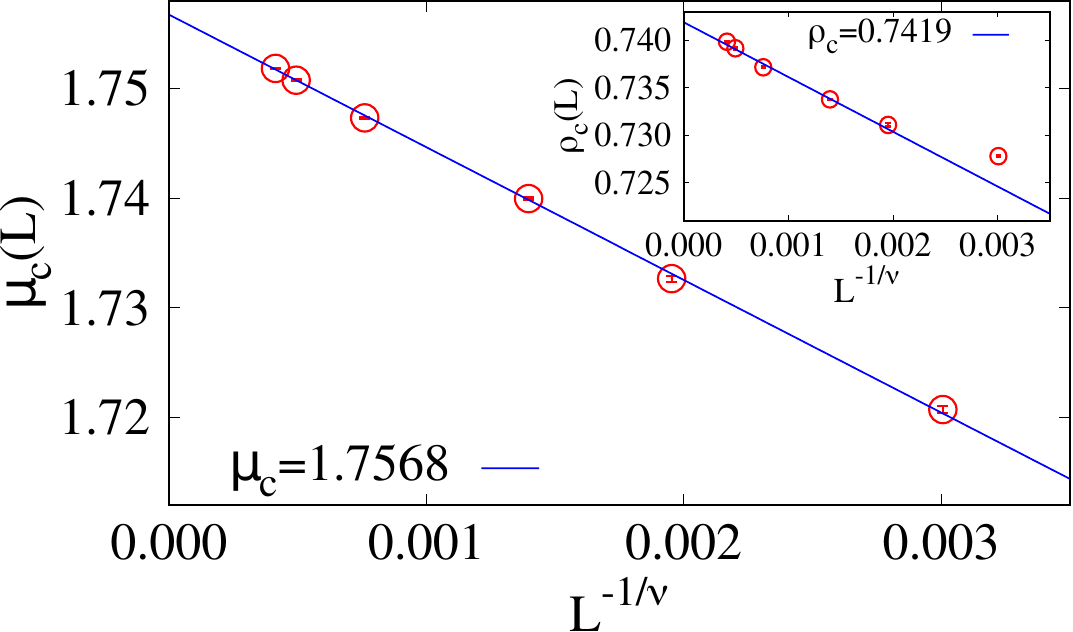}
\caption{\label{fig:muc_fit_2} (Color online)
The extrapolation of $\mu_c(L)$ and $\rho_c(L)$ (inset) to infinite system size using the scaling in Eq.~(\ref{eq:criticalscaling}). The data are for the $2$-NN model and we have chosen  $\nu=2/3$. }
\end{figure}

We determine the  critical chemical potential $\mu_{c}$ and critical density $\rho_c$ by extrapolating  $\mu_c(L)$ and $\rho_c(L)$ to infinite system size using Eq.~(\ref{eq:criticalscaling}).  The extrapolation is shown in  Fig.~\ref{fig:muc_fit_2}. We obtain $\mu_{c}=1.7568(4)$ and $\rho_c = 0.7419(5)$. Our estimate for $\mu_c$ is consistent with earlier estimates (see above). For $\rho_c$, the estimate is closer to the transfer matrix estimate of $0.74856$~\cite{bartelt1984triangular} rather than the more recent Monte Carlo estimate of $0.72(2)$~\cite{zhang2008monte}. From the inset of Fig.~\ref{fig:muc_fit_2}, it can be seen that even for the smallest system size $\rho_c(L) > 0.725$.

\subsection{\label{sec:3nn} $3$-NN Model}

The $3$-NN model has been studied using  matrix methods and series expansion~\cite{1968-OB-JCP-Traingularlattice},  tensor renormalization group method~\cite{akimenko2019tensor}, evaporation-deposition algorithms~\cite{darjani2019liquid} and the SCWL algorithm~\cite{jaleel2021hard}. The model shows a single first order transition from a disordered phase to a sublattice-ordered phase. Based on SCWL algorithm, we recently found that $\mu_c=4.4641(3)$ and critical pressure is $ P_c=0.6397(1) $. Both phases coexist between densities $\rho_{f}= 0.8482(1)$ and $\rho_{s}=0.9839(2)$~\cite{jaleel2021hard}. A detailed discussion of the analysis as well as a summary of  past work can be found in Ref.~\cite{jaleel2021hard}. 
\subsection{\label{sec:4nn} $4$-NN}

Compared to $1$-NN, $2$-NN and  $3$-NN models, there is not much known for $k\geq 4$. The $4$-NN model has been studied earlier using tensor renormalization group~\cite{akimenko2019tensor}. It was concluded that the $4$-NN model shows a  continuous phase transition with $\mu_c$ between  $2.65$ and $2.7$. 

We now present our results for the $4$-NN model in detail. We will follow a similar analysis for larger $k$.

\begin{figure}
	\includegraphics[width=\columnwidth]{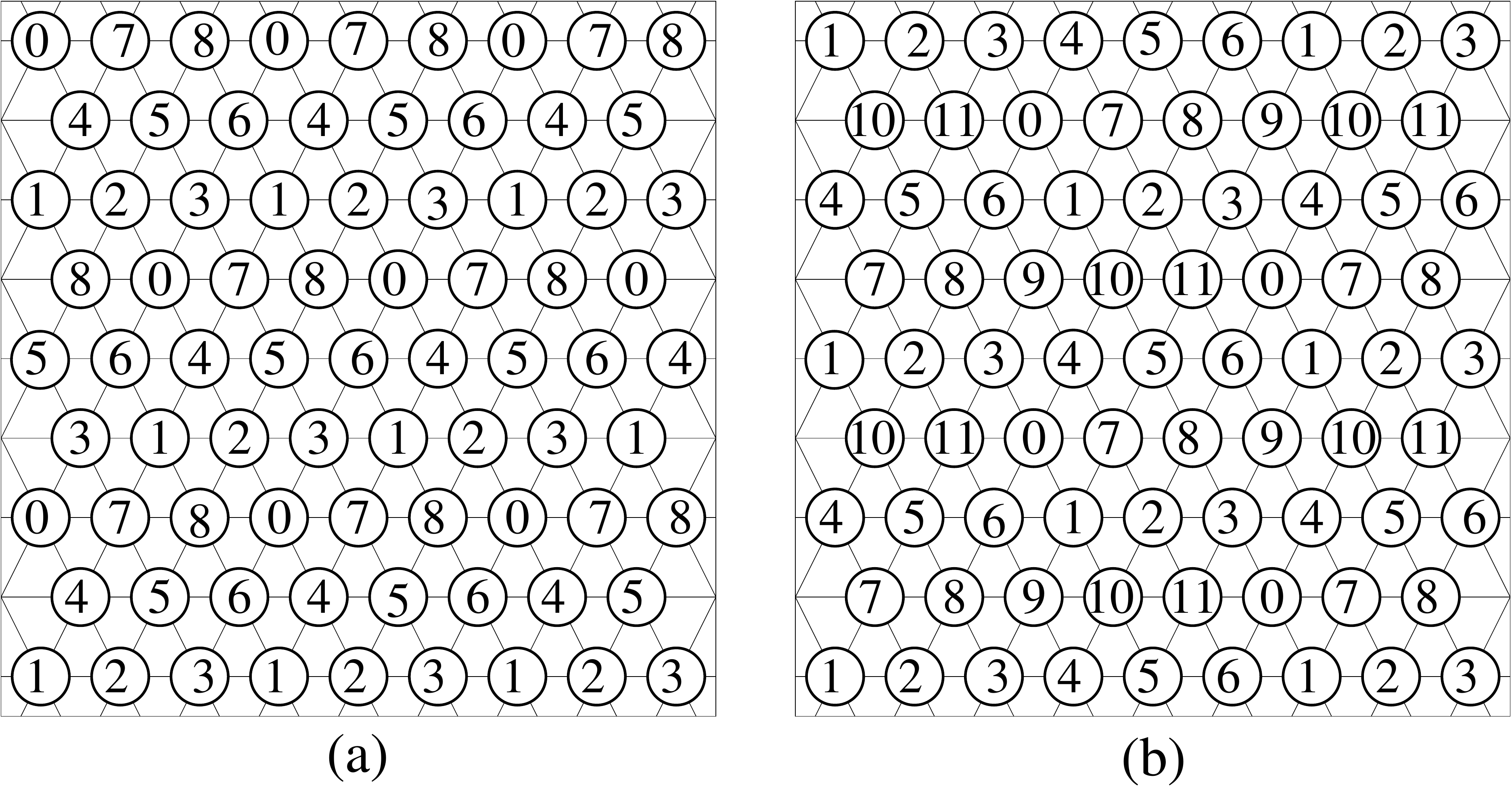}
	\caption{\label{fig:4nnsub} Division of  lattice  sites into sublattices for the (a) $4$-NN model (9 sublattices) and (b) $5$-NN model (12 sublattices).}
\end{figure}

At full packing, the particles occupy one of the $N_{\mathrm{sub}}=9$ sublattices shown in Fig.~\ref{fig:4nnsub}(a). We first show that there is only one phase transition from a disordered fluid phase to the high-density sublattice-ordered phase.

\begin{figure}
	\includegraphics[width=\columnwidth]{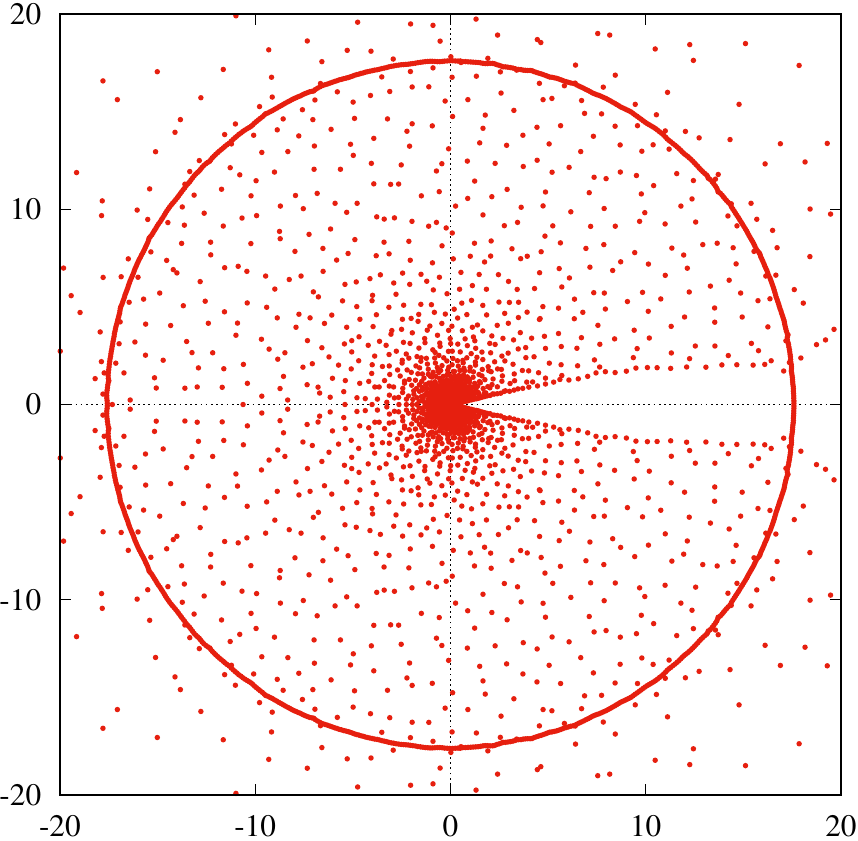}
	\caption{\label{fig:4nnpfz} Zeros of the grand canonical partition function in the complex z-plane
($z=\mathrm{e}^{\mu}$) for the $4$-NN model. The locus of the zeros forms a circle and pinches the positive real axis only at one point. The data are for $L=216$ }
\end{figure}

To do so, we examine the loci of zeros of the the partition function in the complex $z$-plane.  Figure~\ref{fig:4nnpfz} shows the zeros for $L=216$. The partition function zeros (PFZ) pinch the real axis at only one point, showing the presence of only a single phase transition. The locus forms a circle, hence the angle it makes with the real axis is $\pi/2$, suggesting that the  transition is discontinuous~\cite{blythe2003lee,bena2005statistical}. 

\begin{figure}
	\includegraphics[width=\columnwidth]{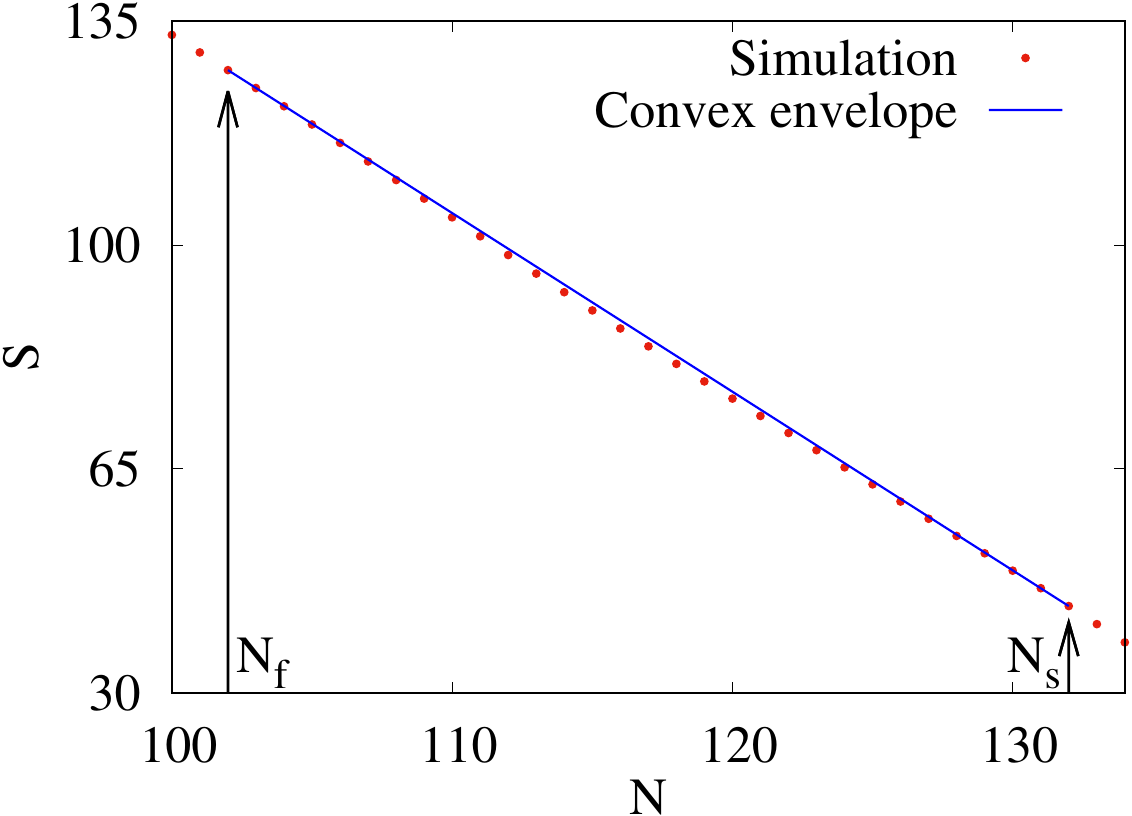}
	\caption{\label{fig:4nnnc}The non-convex part of the entropy and the corresponding convex envelope construction for the $4$-NN model. The  black solid arrows show location of $ N_f $ and $ N_s $. Significance of $ N_f $ and $ N_s $ is explained in text. }
\end{figure}

The first order nature of the transition can be further established by showing that the entropy is non-convex in certain regions~\cite{jaleel2021rejection,jaleel2021hard}. Figure~\ref{fig:4nnnc} shows the non-convex part of the entropy for $L=36$  and the corresponding convex envelope, shown by the solid line. Analysis of the non-convexity  gives precise estimate of critical parameters of the  first order transition~\cite{jaleel2021rejection,jaleel2021hard}. Let $N_f$ and $N_s$ denote  the number of particles at the endpoints  of the convex envelope in the fluid side and sublattice side respectively (see Fig.~\ref{fig:4nnnc}).  Then $\rho_{f}(L)=N_f/L^2\times N_{\mathrm{sub}}$ and $\rho_{s}(L)=N_s/L^2\times N_{sub}$ are the densities at the boundaries of the coexistence regime. The critical chemical potential 
\begin{equation}
\mu_c(L)=-\frac{ S(N_s)-S(N_f)}{N_s-N_f} 
\end{equation}
is the slope of the convex envelope. The system size dependent critical parameters $\rho_f(L)$, $\rho_s(L)$ and $  \mu_c(L)$, thus obtained, are tabulated in Table~\ref{tab:4nntri} for different system sizes.  The convergence to the infinite system size is rapid. 
	\begin{table}		
	\caption{Critical parameters from non-convexity in entropy of $4$-NN model. The last row shows the results of extrapolation to infinite $L$ using Eq.~(\ref{eq:criticalscaling}) with $\nu=1/2$  (see Fig.~\ref{fig:nc4nn}).}
	\begin{ruledtabular}
		\begin{tabular}{cccc}
			$L$& $\rho_{f}$ &$\rho_{s}$ & $ \mu_c $   \\
			\hline		
108 & 0.7319(1)&0.9094(1)&2.8593(1)\\ 
117 & 0.7331(1)&0.9090(1)&2.8606(1)\\ 
126 & 0.7340(1)&0.9089(1)&2.8632(1)\\ 
135 & 0.7347(1)&0.9085(1)&2.8632(1)\\ 
144 & 0.7356(1)&0.9086(1)&2.8643(1)\\ 
153 & 0.7358(1)&0.9081(1)&2.8644(1) \\ 
162 & 0.7364(1)&0.9080(1)&2.8651(1)\\  
171 & 0.7366(1)&0.9077(1)&2.8655(1) \\  
180 & 0.7374(1)&0.9077(1)&2.8661(1)\\ 
189 &  0.7376(1)&0.9076(1)&2.8663(1)\\ 
198 &  0.7379(1)&0.9076(1)&2.8666(1)\\ 
207 &  0.7382(1)&0.9075(1)&2.8669(1) \\ 
216 &  0.7384(1)&0.9072(1)&2.8671(1)\\ 
225 &   0.7386(1)&0.9073(1)&2.8675(1) \\   	
$\infty$ & 0.7404(1)&0.9067(1)&2.8696(1) \\			
		\end{tabular}
	\end{ruledtabular}
	\label{tab:4nntri}
\end{table}

\begin{figure}
	\includegraphics[width=\columnwidth]{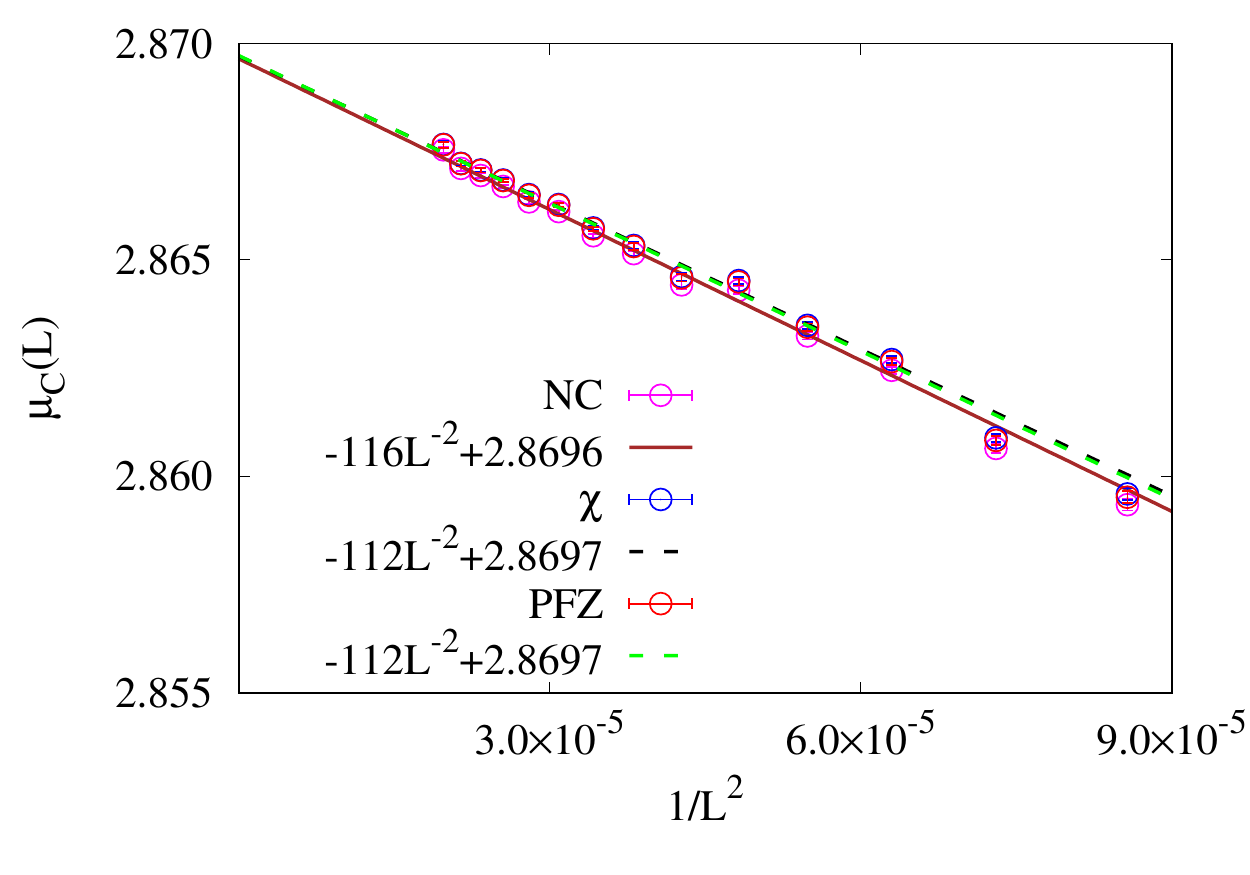}
	\caption{\label{fig:nc4nn} Variation of the critical chemical potential  $\mu_c(L)$, with system size L for $4$-NN model. The data are obtained from non-convexity (NC) analysis, susceptibility $\chi$ and partition function zeros (PFZ). The error in each data point was obtained from 16 independent simulations. Solid and dashed lines are the best linear fit to the data. }
\end{figure}

Figure~\ref{fig:nc4nn} shows the extrapolation of $\mu_c(L)$ to infinite system size using Eq.~(\ref{eq:criticalscaling}) with $\nu=2$. We determine $\mu_c(L)$ from the maximum of $\chi$, NC analysis and the position of the PFZ closest to the origin. We obtain  $\mu_{c,4}=2.8696(2)$ from NC analysis, and $\mu_{c,4}=2.8697(2)$ from susceptibility and PFZ. The coexistence densities are found to $\rho_{f,4}=0.7404(2)$ and $\rho_{s,4}=0.9067(2)$ from NC (extrapolation not shown). The critical pressure is estimated to be  $P_{c,4}=0.3262(2)$. 

Further results for the dependence of  density,  order parameter,  Binder cumulant, and pressure with chemical potential $\mu$ can be found in the Supplemental Material~\cite{supp} under section $4$-NN.  

Note that our analysis shows that the transition is first order. This is in contradiction to that obtained from TRG method where it was concluded that the transition is continuous. Also, our estimate of $\mu_c =2.8696(2) $ is outside the range of $2.65$ and $2.7$
estimated using TRG in Ref.~\cite{akimenko2019tensor}. The better accuracy is presumably due to the SCWL algorithm accessing the high-density states in an efficient manner.

\subsection{\label{sec:5nn} $5$-NN Model}

In the $5$-NN model, at full packing, the particles occupy one of the  $N_{\mathrm{sub}}=12$ sublattices shown in Fig.~\ref{fig:4nnsub}(b). In earlier work it was argued, based on the tensor renormalization group, that the model exhibits a single first order transition with $\mu_c$ roughly in the range $ 4.4 < \mu_{c} < 4.5 $~\cite{akimenko2019tensor}.

\begin{figure}
	\includegraphics[width=\columnwidth]{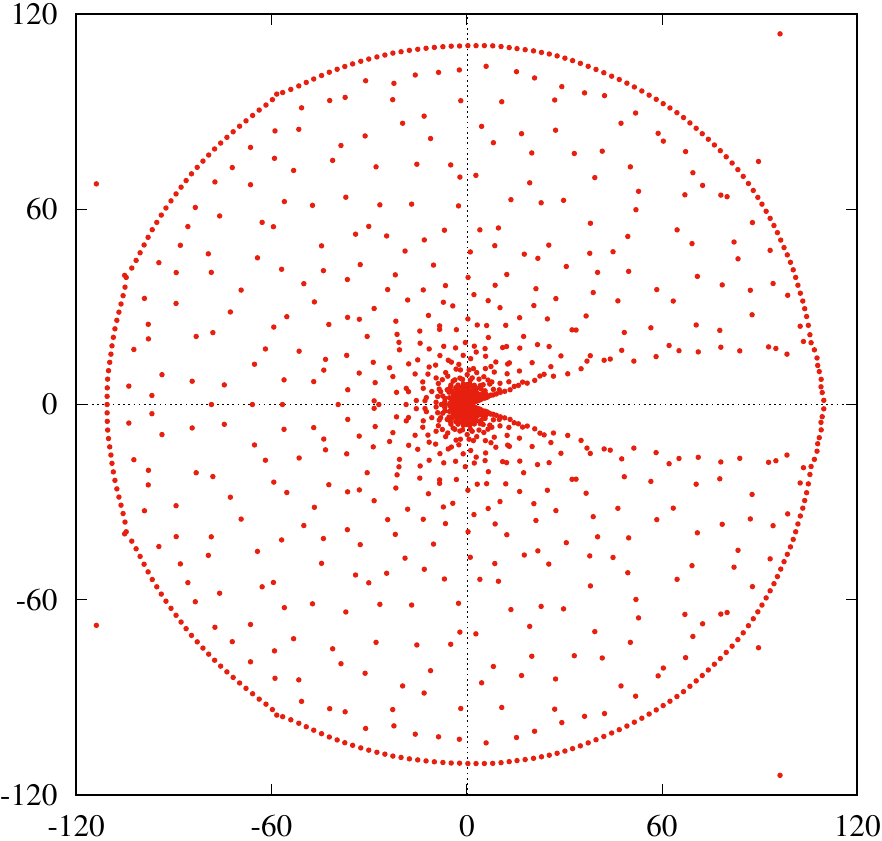}
	\caption{\label{fig:5nnpfz} Zeros of the grand canonical partition function in the complex z-plane
		($z=\mathrm{e}^{\mu}$ ) for $5$-NN model. The locus of the zeros forms a circle and
		pinches the positive real axis only at one point.}
\end{figure}

The analysis we present for the $5$-NN model  (and higher $k$) is same as that of the $4$-NN model, and we only summarize the results. We simulate system sizes up to $L=180$. We first show that there is only one phase transition from a disordered fluid phase to the high-density sublattice-ordered phase.
Figure~\ref{fig:5nnpfz}  shows the PFZ for $L=180$. The locus of the zeros form a circle and pinch the real axis at only one point. We conclude that there is only one phase transition in the system. Since the locus is a circle, the angle of approach of the leading zeros is $\pi/2$, suggesting that the transition is discontinuous.  We note that for smaller $L$, the  PFZ of $5$-NN model also forms an inner circle. However, as $L$ increases, these two circles merge  and become indistinguishable. 

\begin{figure}
	\includegraphics[width=\columnwidth]{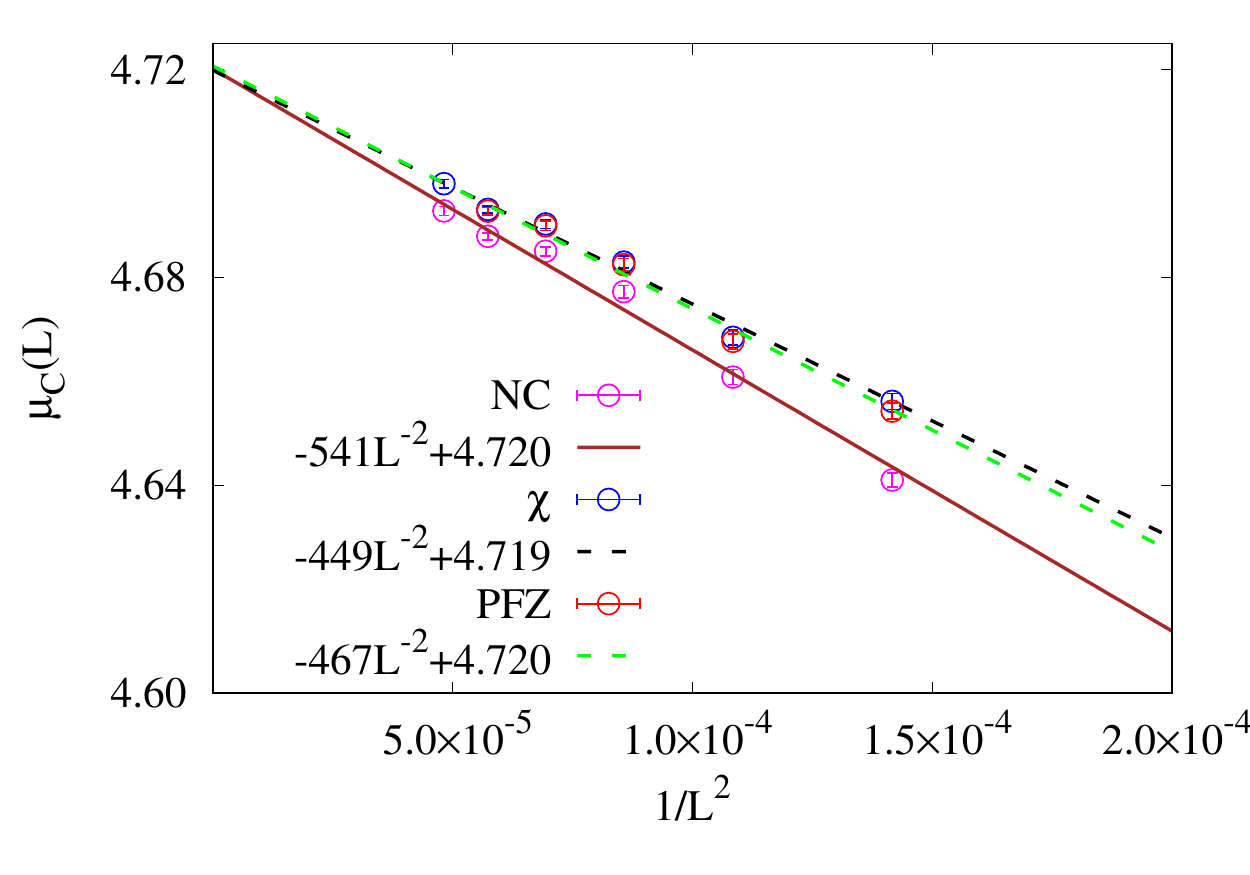}
	\caption{\label{fig:nc5nn}  Variation of the critical chemical potential  $\mu_c(L)$, with system size L for $5$-NN model.
The data are obtained from non-convexity (NC) analysis, susceptibility $\chi$ and partition function zeros (PFZ). The
error in each data point was obtained from 16 independent simulations. Solid and
dashed lines are the best linear fit to the data. }
\end{figure}

We find that the entropy is non-convex, giving  direct proof of the first order nature of the transition.  From the NC analysis, we obtain $\rho_f(L)$, $\rho_s(L)$, and $  \mu_c(L)$ for different system sizes.  Figure~\ref{fig:nc5nn} shows the extrapolation to infinite system size of  $\mu_c(L)$ obtained from non-convexity analysis, PFZ and  the peak position of susceptibility. The critical chemical potential was estimated to be  $\mu_{c,5}=4.720(3)$ from NC analysis and PFZ data, and $\mu_{c,5}=4.720(1)$ from susceptibility. The endpoints of the coexistence regime were similarly found to be $\rho_{f,5}=0.916(3)$ and $\rho_{s,5}=0.988(3)$  from the NC  analysis. We obtain the  critical pressure to be $P_{c,5}=0.3942(2)$.

Further results for the dependence of  density,  order parameter,  Binder cumulant, and pressure with chemical potential $\mu$ can be found in the Supplemental Material~\cite{supp} under section $5$-NN.  

Note that our analysis shows that the transition is first order, consistent with preliminary results obtained using TRG~\cite{akimenko2019tensor}. Also, our estimate of $\mu_c =4.720(3) $ is outside the range of $ 4.4 < \mu_{c} < 4.5 $ obtained  in Ref.~\cite{akimenko2019tensor}. 

\subsection{\label{sec:6nn} $6$-NN Model}

\begin{figure}
	\includegraphics[width=\columnwidth]{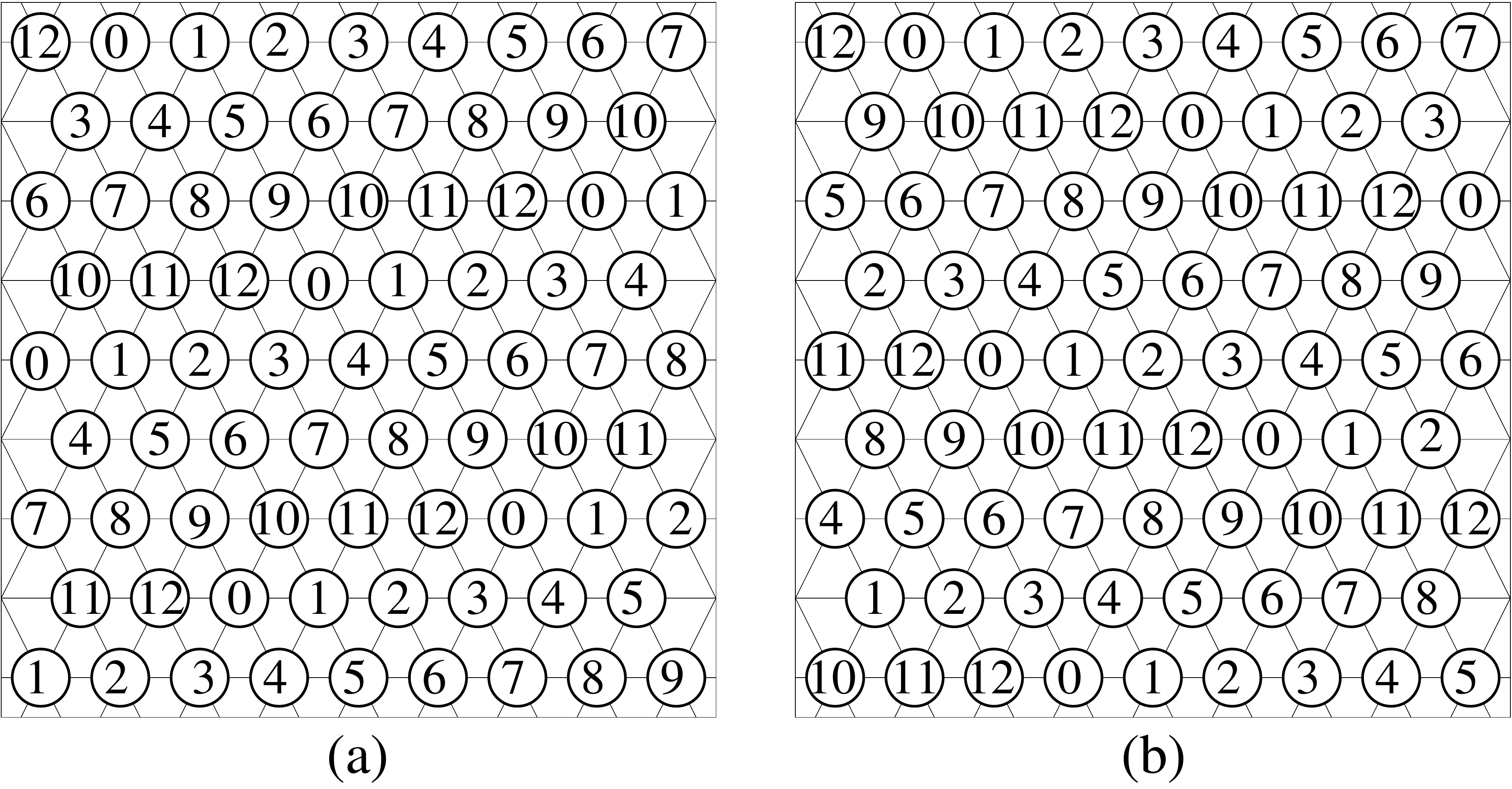}
	\caption{\label{fig:6nnsub} Division of  lattice  sites into sublattices for the $6$-NN model (13 sublattices).  The division can be done in two equivalent ways as shown in (a) and (b)}
\end{figure}

In the $6$-NN model, at full packing, the particles occupy one of the  $N_{\mathrm{sub}}=13$ sublattices shown in Fig.~\ref{fig:6nnsub}. The lattice can be divided into 13 sublattices in two different ways, which we call as  type-A and type-B, as shown in Fig.~\ref{fig:6nnsub}. The nature and location of  the transition is not known for this model as it has not been studied earlier.  

\begin{figure}
	\includegraphics[width=\columnwidth]{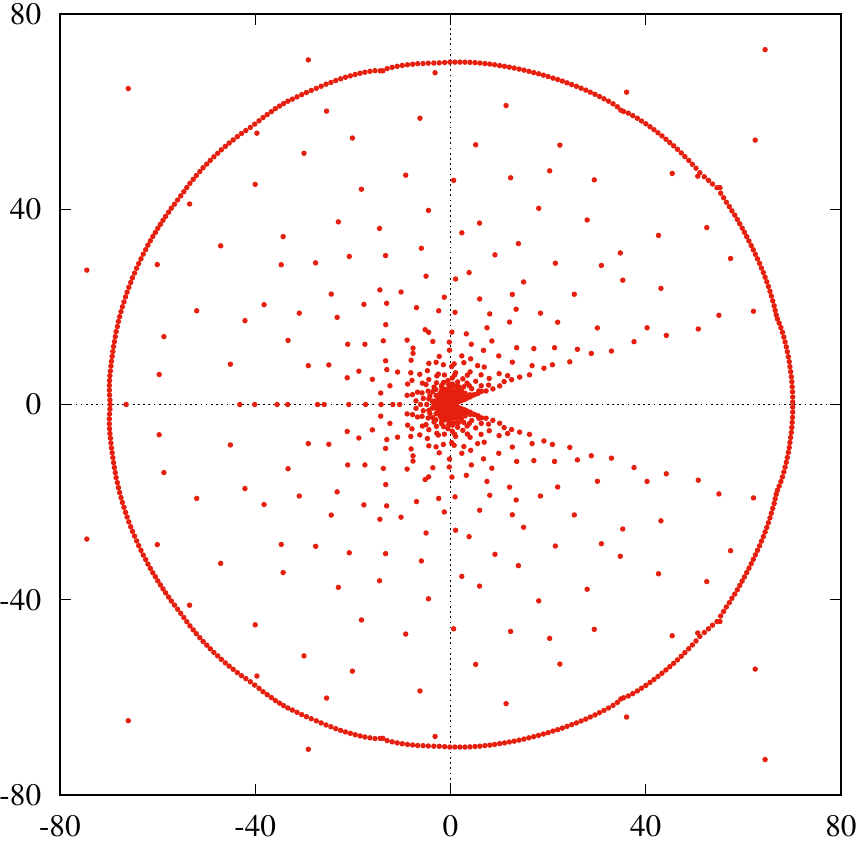}
	\caption{\label{fig:6nnpfz} Zeros of the grand canonical partition function in the complex z-plane
		($z=\mathrm{e}^{\mu}$ ) for $6$-NN model. The locus of the zeros forms a circle and
		pinches the positive real axis only at one point.}
\end{figure}

We simulate system sizes upto $L=169$. We first show that there is only one phase transition from a disordered fluid phase to the high-density sublattice-ordered phase.
In Fig.~\ref{fig:6nnpfz} we show the PFZ for $L=169$. The locus of the zeros form a circle and pinch the real axis at only one point. We conclude that there is only one phase transition in the system. Since the locus is a circle, the angle of approach of the leading zeros is $\pi/2$, suggesting that the transition is discontinuous.  

\begin{figure}
	\includegraphics[width=\columnwidth]{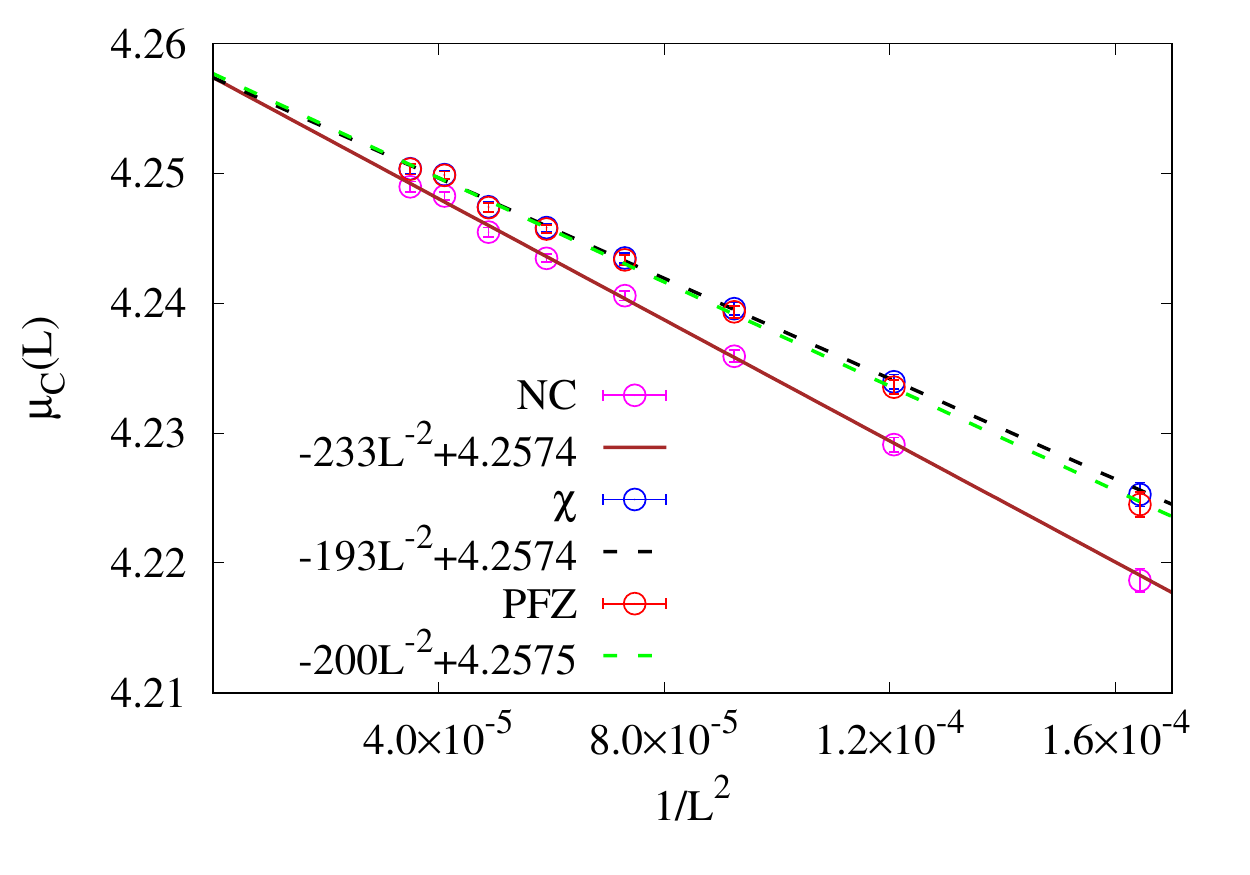}
	\caption{\label{fig:nc6nn} Variation of the critical chemical potential  $\mu_c(L)$, with system size L for $6$-NN model. The data are obtained from non-convexity (NC) analysis, susceptibility $\chi$ and partition function zeros (PFZ). The error in each data point was obtained from 16 independent simulations. Solid and dashed lines are the best linear fit to the data. }
\end{figure}

We find that the entropy is non-convex, giving a direct proof of the first order nature of the transition.   From the NC analysis, we obtain $\rho_f(L)$, $\rho_s(L)$, and $  \mu_c(L)$ for different system sizes.  Figure~\ref{fig:nc6nn} shows the extrapolation to infinite system size of  $\mu_c(L)$ obtained from non-convexity analysis, PFZ and  the peak position of susceptibility. The critical chemical potential is estimated to be  $\mu_{c,6}=4.2574(4)$ from non-convexity analysis and  susceptibility and  $\mu_{c,6}=4.2575(4)$ from PFZ data. The endpoints of the coexistence regime were similarly found to be $\rho_{f,6}=0.7898(1)$ and $\rho_{s,6}=0.9818(2)$  from the NC  analysis. We obtain the  critical pressure to be $P_{c,6}=0.3287(1)$.

Further results for the dependence of  density,  order parameter,  Binder cumulant, and pressure with chemical potential $\mu$ can be found in the Supplemental Material~\cite{supp} under section $6$-NN.

\subsection{\label{sec:7nn} $7$-NN Model}

In the $7$-NN model, at full packing, the particles occupy one of the  $N_{\mathrm{sub}}=16$ sublattices shown in Fig.~\ref{fig:4nnsub}(b). The nature and location of transition are not known for this model as it is not studied earlier. 
  \begin{figure}
  	\includegraphics[width=0.7\columnwidth]{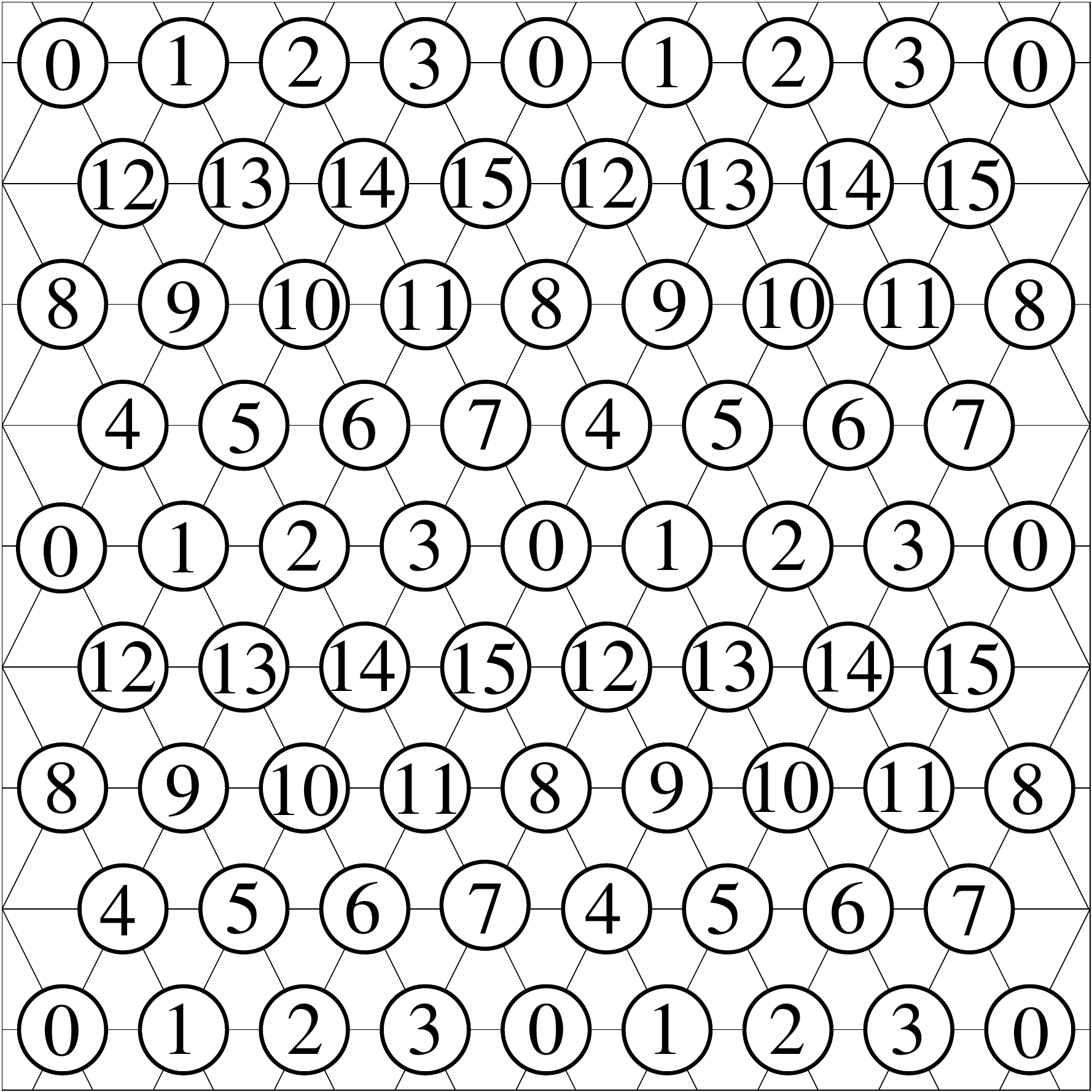}
  	\caption{\label{fig:7nnsub} Division of  lattice  sites into sublattices for the $7$-NN model (16 sublattices). }
  \end{figure}
  
\begin{figure}
	\includegraphics[width=\columnwidth]{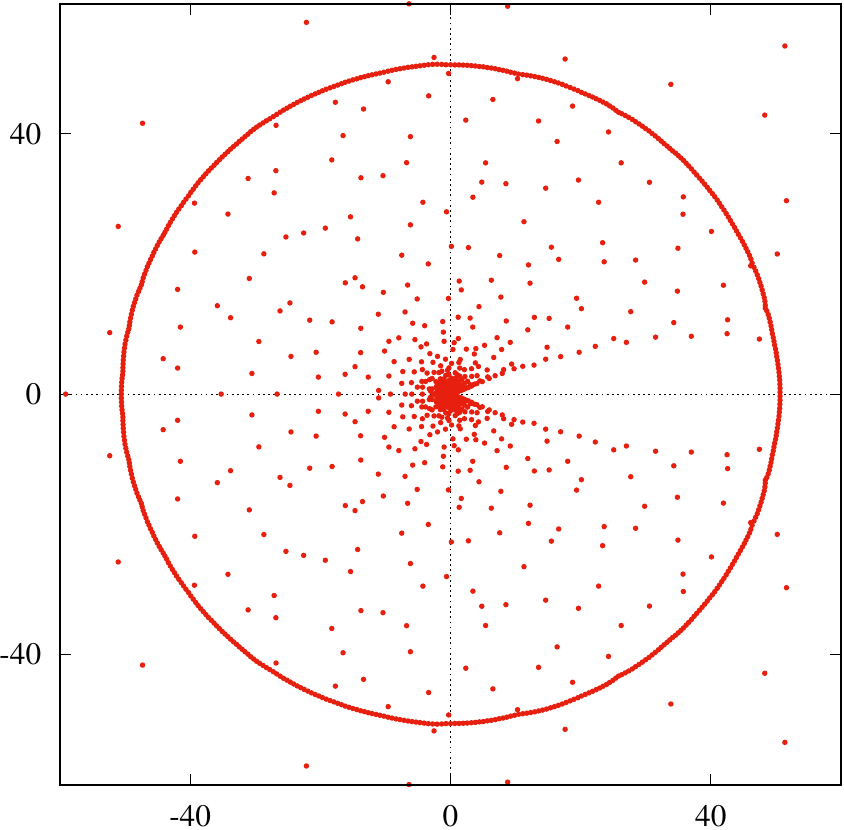}
	\caption{\label{fig:7nnpfz} Zeros of the grand canonical partition function in the complex z-plane
		($z=\mathrm{e}^{\mu}$ ) for $7$-NN model. The locus of the zeros forms a circle and
		pinches the positive real axis only at one point.}
\end{figure}

We simulate system sizes up to $L=192$. We first show that there is only one phase transition from a disordered fluid phase to the high-density sublattice-ordered phase.
In Fig.~\ref{fig:7nnpfz} we show the PFZ for $L=192$. The locus of the zeros form a circle and pinch the real axis at only one point. We conclude that there is only one phase transition in the system. Since the locus is a circle, the angle of approach of the leading zeros is $\pi/2$, suggesting that the transition is discontinuous.  

\begin{figure}
	\includegraphics[width=\columnwidth]{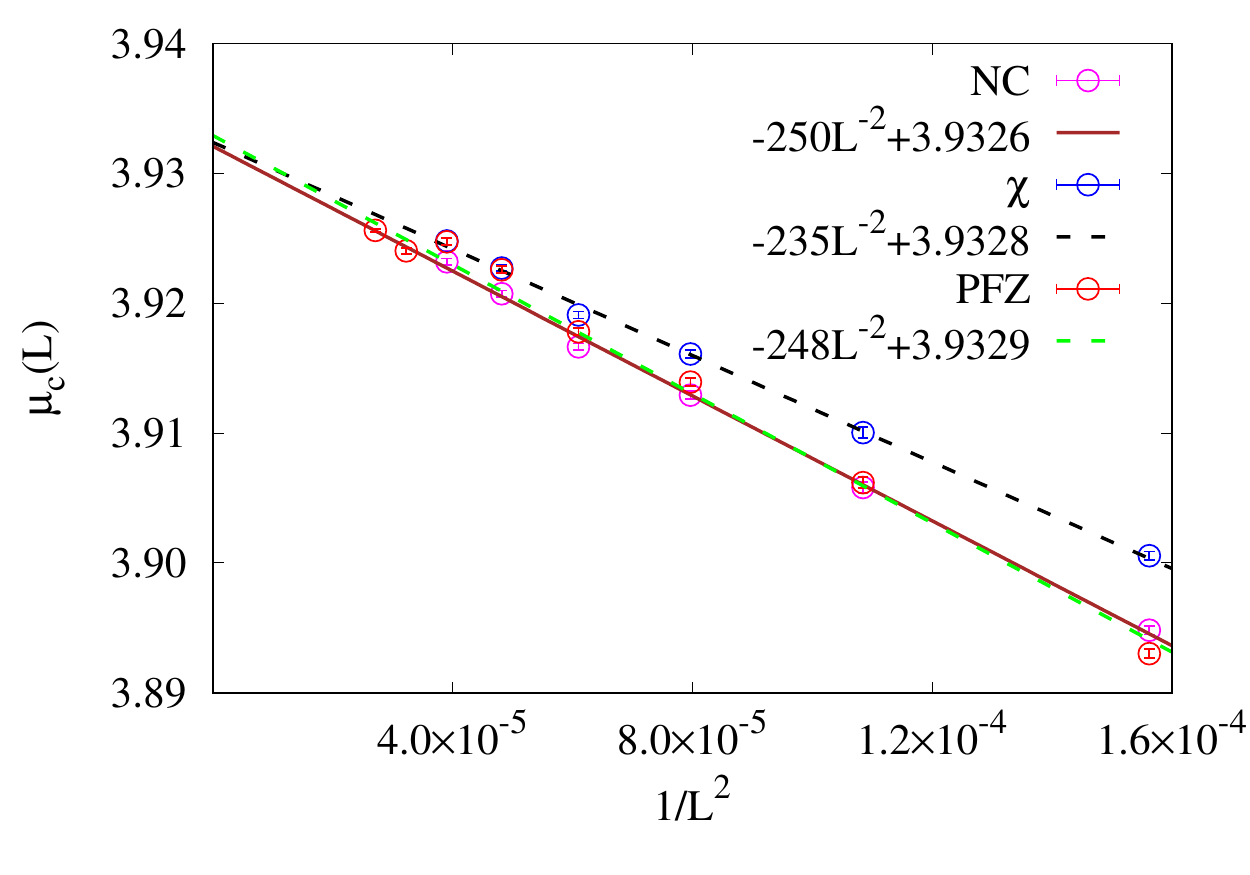}
	\caption{\label{fig:nc7nn} Variation of the critical chemical potential  $\mu_c(L)$, with system size L for $7$-NN model. The data are obtained from non-convexity (NC) analysis, susceptibility $\chi$ and partition function zeros (PFZ). The error in each data point was obtained from 16 independent simulations. Solid and dashed lines are the best linear fit to the data.  }
\end{figure}

We find that the entropy is non-convex, giving a direct proof of the first order nature of the transition.  From the NC analysis, we obtain $\rho_f(L)$, $\rho_s(L)$, and $  \mu_c(L)$ for different system sizes.  Figure~\ref{fig:nc7nn} shows the extrapolation to infinite system size of  $\mu_c(L)$ obtained from non-convexity analysis, PFZ and  the peak position of susceptibility. The critical chemical potential was estimated to be  $\mu_{c,7}=3.9326(8)$ from non-convexity analysis and $\mu_{c,7}=3.9328(8)$ from susceptibility and $\mu_{c,7}=3.9329(7)$ from PFZ data. The endpoints of the coexistence regime were similarly found to be $\rho_{f,7}=0.7469(4)$ and $\rho_{s,7}=0.9709(3)$  from the NC  analysis. We obtain the  critical pressure to be $P_{c,7}=0.2471(1)$.

Further results for the dependence of  density,  order parameter,  Binder cumulant, and pressure with chemical potential $\mu$ can be found in the Supplemental Material~\cite{supp} under section $7$-NN.

\section{\label{sec:summary}Summary and Discussion }

In this paper, we studied the phase transitions in the $k$-NN hard core lattice gas model on the triangular lattice. In this model, a particle excludes all the  sites up to the $k$-th next-nearest neighbors from being occupied by another particle. We obtain the phase diagram and quantify the critical behavior for $k\leq 7$ using the SCWL flat histogram algorithm, which incorporates rejection free cluster moves to evolve the system.

\begin{figure}
	\includegraphics[width=\columnwidth]{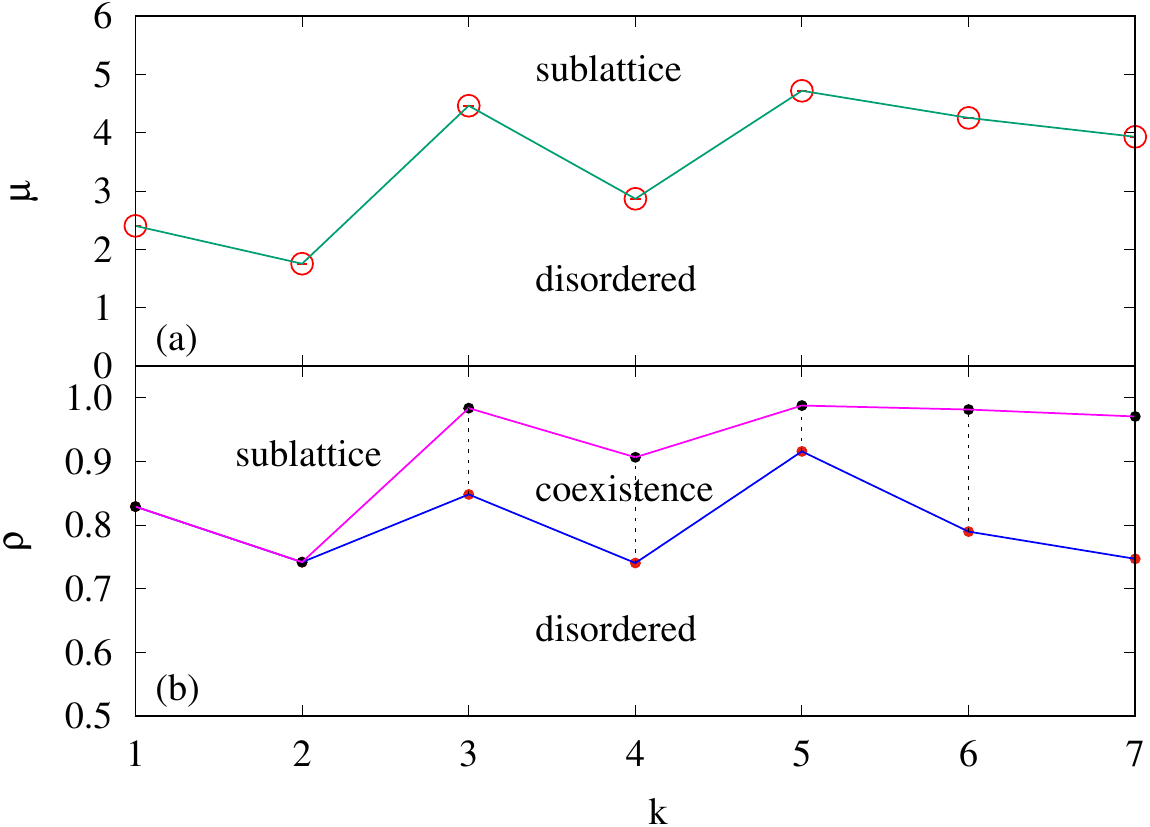}
	\caption{\label{fig:mrknn} The phase diagram of the $k$-NN model in the (a) $\mu$-$k$ plane and (b) $\rho$-$k$ plane. }
\end{figure}

We use the $1$-NN model or the hard hexagon model, for which an exact solution is known, to benchmark our simulations. For the $2$-NN model, we obtain better estimates for the critical density. We show that  the $4$-NN to $7$-NN models undergo only a single first order phase transition from a low-density disordered phase to a high-density sublattice-ordered phase. For each of these models, we obtain accurate estimates of the critical chemical potential, the densities of the disordered  and sublattice phases in the coexistence regime, and the critical pressure. The quantitative results are summarized in  Table~\ref{tab:results}. The phase diagram in the $\mu$-$k$ plane and the $\rho$-$k$ plane are shown in Fig.~\ref{fig:mrknn}. Though $\mu_c$ on an average increases with $k$ [see Fig.~\ref{fig:mrknn}(a)], we are unable to quantify the increase based only on results for $k\leq 7$. For the coexistence densities, with increasing $k$, we observe that $\rho_s$ tends to $1$ while $\rho_f$ saturates around $0.8$ [see Fig.~\ref{fig:mrknn}(b)]. 
\begin{table}		
	\caption{\label{tab:results}Summary of the coexistence densities of the disordered phase ($\rho_f$), the sublattice phase ($\rho_s$), the critical chemical potential ($\mu_c$) and critical pressure ($P_c$) for different $k$.  For $k=1$, the results are exact~\cite{1980-b-jpa-exact,baxterBook}, and the results for $k=3$ are taken from Ref.~\cite{jaleel2021hard}. }
	\begin{ruledtabular}
		\begin{tabular}{lllll}
			$ 	k $ &$\rho_{f}$ &$\rho_{s}$ & $ \mu_c$   & $ P_c $ \\
			\hline
			
			1 &  -   & 0.82917   &   2.4060&
			\\
			2  &  -   & 0.7419(5)    &       1.7568(4)&
			\\	
			3 &  0.8482(1)    &   0.9839(2) &       4.4641(3)&0.6397(1)
			\\
			4 & 0.7404(2)&0.9067(2)&2.8696(2)&0.3262(2)
			\\
			5 & 0.916(3)   &    0.988(3) &       4.720(3)& 0.3942(2)
			\\
			6 &  0.7898(1) &    0.9818(2) &       4.2574(4)&0.3287(1) \\	
			7  & 0.7469(4)   &    0.9709(3)  &       3.9315(6)&0.2471(1)	
		\end{tabular}
	\end{ruledtabular}
\end{table}

The results obtained in the paper allow us to compare the efficacies of the TRG method and SCWL algorithm. 
The TRG method has been used to study the $k$-NN model on the triangular lattice~\cite{akimenko2019tensor}. The method yields good results for $k=1,2,3$. However,  the results for $4$-NN and $5$-NN differ considerably from the results that we obtain using SCWL. For the $4$-NN model, TRG method predicts a continuous transition. However, we show clearly that the transition is first order. In addition, the extrapolated critical chemical potential obtained in this paper does not lie in the range predicted by TRG both for $k=4$ and $k=5$. It would be of interest to see how the TRG method can be improved to be able to study HCLG of particles with large excluded volume.

For infinite $k$, the $k$-NN model resembles the hard sphere model for which an intermediate hexatic phase separates a fluid and orientationally ordered high-density phase~\cite{1973-kt-jpc-ordering,1979-y-prb-melting,1979-nh-prb-dislocation,2011-bk-prl-twostep,engel2013hard,kapfer2015two}. Thus, one would expect that signatures of these continuum phases are seen in the $k$-NN model for large enough $k$. For the $k$ that we have studied, there is only a single phase transition, and it does not appear that the continuum limit can be obtained for a $k$ that is computationally tractable.

The SCWL algorithm, as seen in this paper, appears well-suited to study phase transitions in HCLG models.  The power of the flat histogram method is even more when applied to density of states which depend on more than one variable. A viable example is the binary gas, the simplest of which is the mixture of $1$-NN and $0$-NN particles which shows a non-trivial phase diagram with a tri-critical point~\cite{1983-p-jcp-coexistence,2015-os-pre-transfer,liu2001phase,rodrigues2019three,rodrigues2019thermodynamic, rodrigues2020fluid}. Similarly, multi-dimensional density of states appear in HCLG with additional attractive interactions~\cite{orban1968jcp,prestipino2022condensation}. These are promising areas for future study.

\end{document}


|
\title{\textit{Supplemental material for} \\\textbf{The freezing phase transition in hard core lattice gases on  triangular lattice with exclusion up to seventh next nearest neighbor} }
\author{\textbf{Asweel Ahmed A Jaleel$^{1,2,3}$, Dipanjan Mandal$^4$, Jetin E. Thomas$^{1,2}$ , R. Rajesh$^{1,2}$} \\
\textit{$^1$The Institute of Mathematical Sciences, C.I.T. Campus,
	Taramani, Chennai 600113, India} \\
\textit{$^2$Homi Bhabha National Institute, Training School Complex, Anushakti Nagar, Mumbai 400094, India} \\\textit{ $^3$ Department of Physics, Sadakathullah Appa College, Tirunelveli, Tamil Nadu 627011, India}
\\\textit{ $^4$ Department of Physics, University of Warwick, Coventry CV4 7AL, United Kingdom}}
\date{\vspace{-5ex}}
\maketitle
Here, we provide further evidence for the first order nature of phase transition from disordered phase to the sublattice-ordered phase in the $4$-NN to $7$-NN  models on triangular lattice.  The thermodynamic quantities  pressure $ P $ and $\widetilde{P}$, order parameter $q_k$ and susceptibility $\chi$ are defined in Eqs. (9) - (14)  of the paper.  The Binder cumulant $ u_k $ associated with $q_k$ is defined as
\begin{equation}
u_k = 1- \frac{\langle q_k^4 \rangle}{2\langle q_k^2 \rangle^2} , 
\end{equation} 
for $k=4,5 $ and 7
\section*{$4$-NN model}
\begin{figure}
\centering
\includegraphics[width=\textwidth]{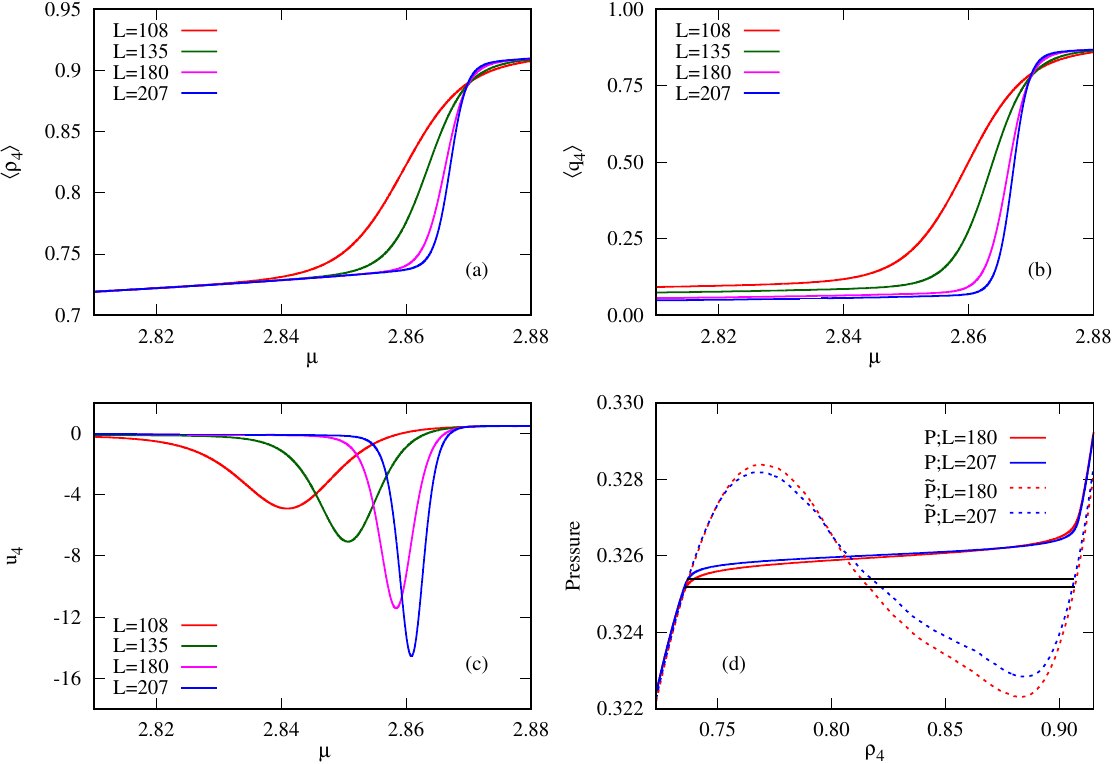}
\caption{ \label{S1} The variation of (a) average density $\langle \rho_4 \rangle$,   (b) average order parameter $\langle q_4 \rangle$ and (c) Binder cumulant $u_4 $  as a function of chemical potential $\mu$ for $4$-NN model. (d) The variation of pressure $ P $ calculated from  Eq.~(9) of paper and $\widetilde{P}$ calculated from  Eq.~(11) of paper. The  black solid lines are constant pressure lines estimated from Maxwell's equal area construction to $\widetilde{P}$. Bottom line corresponds to $ L=180 $ and top line corresponds to $  L=207 $. 
}
\end{figure}

In Fig.~\ref{S1} we  show the variation of average density $\langle \rho_4 \rangle$, average order parameter $\langle q_4 \rangle$ and  Binder cumulant $u_4$ with  chemical potential $\mu$ and pressure with density. The density rises sharply around $\mu \approx 2.87$ with the discontinuity becoming sharper with increasing system size, consistent with a first order transition. To show that the  transition is between disordered phase and sublattice-ordered phase, we have plotted order parameter in Fig.~\ref{S1}(b). The order parameter $\langle q_4 \rangle$ is close to zero for disordered phase because all the sublattice densities are on average equal. $\langle q_4 \rangle $  is close to 1 after $\mu \approx 2.87$ implying sublattice-ordered phase where one of the 9 sublattice is predominantly occupied.

The diverging negative peaks in Binder cumulant validates the first order nature of transition. As a further proof for first order nature of transition, we have plotted pressure in Fig.~\ref{S1}(d). The  pressure loops in $\widetilde{P}$ and flat curves in $P$ are characteristics of first order transition. From pressure loops in $\widetilde{P}$, we can estimate constant pressure curves using maxwell's equal area rule. With increase in the system size $ L $, constant pressure lines move closer to grand canonical estimate of pressure $ P $.  

\section*{$5$-NN model}

\begin{figure}
	\centering
	\includegraphics[width=\textwidth]{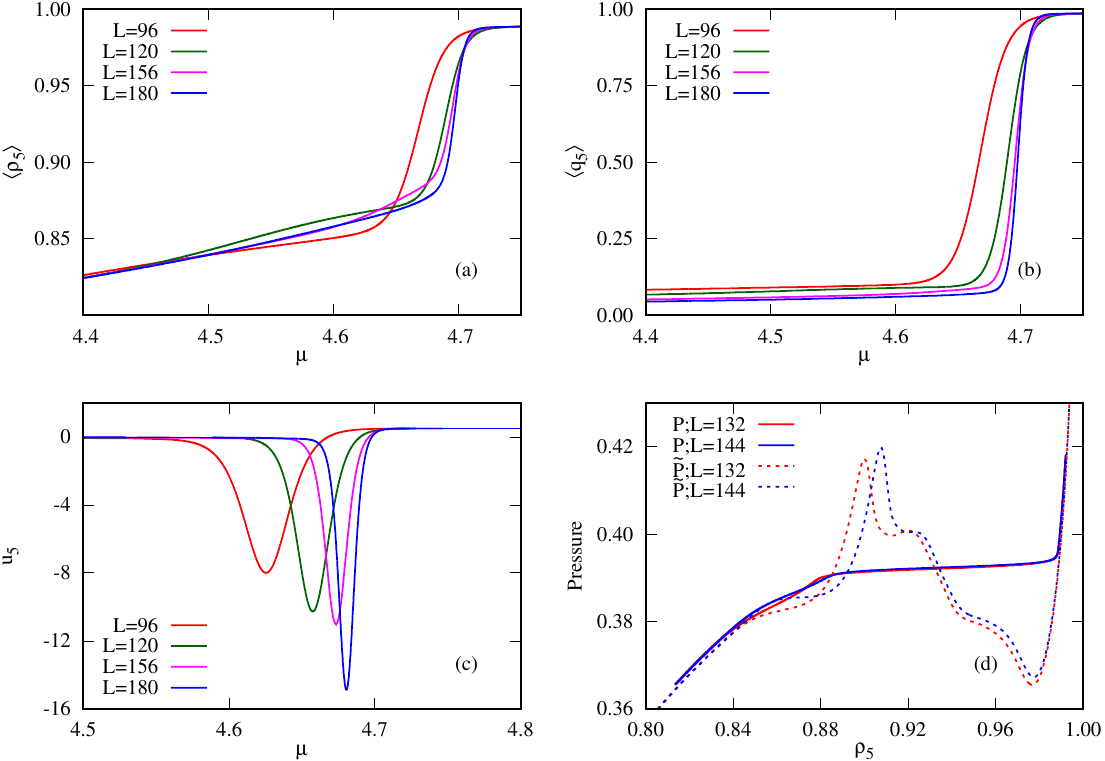}
	\caption{ \label{S2} The variation of (a) average density $\langle \rho_5 \rangle$,   (b) average order parameter $\langle q_5 \rangle$ and (c) Binder cumulant $u_5 $  as a function of chemical potential $\mu$ for $5$-NN model. (d) The variation of pressure $ P $ calculated from  Eq.~(9) of paper and $\widetilde{P}$ calculated from  Eq.~(11) of paper.   }
\end{figure}

In Fig.~\ref{S2}(a), we plotted average density $\langle \rho_5 \rangle$ against $\mu$. The density rises sharply around $\mu \approx 4.72$ with the discontinuity becoming sharper with increasing system size, consistent with a first order transition. To show that transition is between disordered phase and sublattice-ordered phase, we have plotted order parameter in Fig.~\ref{S2}(b). In the disordered phase, all $12$ sublattice densities are on an average  equal and hence $\langle q_5 \rangle=0$ as seen in Fig.~\ref{S2}(b). In the sublattice phase one of the sublattice is predominantly occupied and  $\langle q_5 \rangle $  is close to 1 after $\mu \approx 4.72$ implies sublattice-ordered phase.

The diverging negative peaks in Binder cumulant validates the first order nature of transition.  
As a further evidence for first order nature of transition, we have plotted pressure in Fig.~\ref{S4}(d). The pressure loops in $\widetilde{P}$ and  flat curves in $P$ are characteristics of first order transition.

\section*{$6$-NN model}
\begin{figure}
	\includegraphics[width=\textwidth]{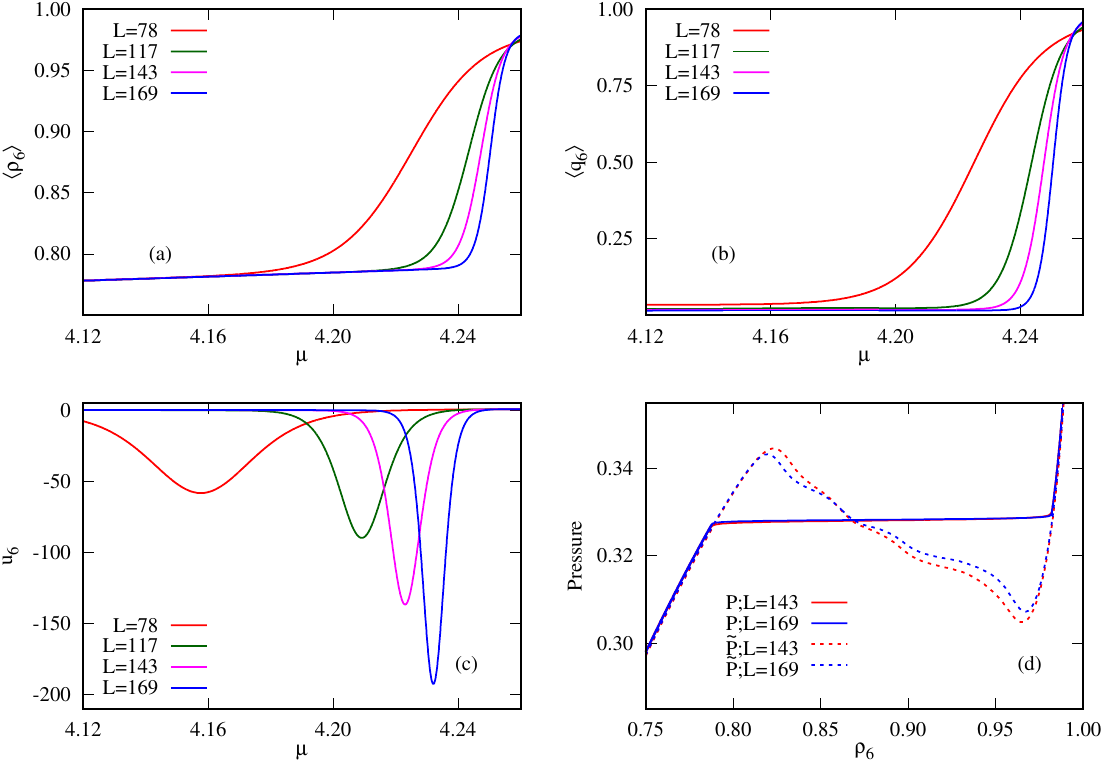}
	\caption{\label{S3} The variation of (a) average density $\langle \rho_6 \rangle$,   (b) average order parameter $\langle q_6 \rangle$ and (c) Binder cumulant $u_6 $  as a function of chemical potential $\mu$ for $6$-NN model. (d) The variation of pressure $ P $ calculated from  Eq.~(9) of paper and $\widetilde{P}$ calculated from  Eq.~(11) of paper.  
	}
\end{figure}

In Fig.~\ref{S3}(a), we  plotted average density $\langle \rho_6 \rangle$ against $\mu$. The density rises sharply around $\mu \approx 4.26$ with the discontinuity becoming sharper with increasing system size, consistent with a first order transition.  To show that transition is between disordered phase and sublattice-ordered phase, we have plotted order parameter in Fig.~\ref{S3}(b). Since there are two ways of defining sublattices, order parameter for $6$-NN model is slightly different from other models. Let us define
\bea
Q_A =\sum_{j=1}^{N_{sub}} \rho_j^A\exp \left[2\pi i\frac{ (j-1)}{N_{sub}}\right],\\
Q_B =\sum_{j=1}^{N_{sub}} \rho_j^B\exp \left[2\pi i\frac{ (j-1)}{N_{sub}}\right],
\eea
where $\rho_j^X$ is the particle density on  $j$-th sublattice for X-type  sublattice division.
Non-zero $Q_A$ or $Q_B$ implies that a particular type of sublattice is occupied preferentially. The order parameter $q_6$ is defined to be
\be
q_6=||Q_A|-|Q_B||. \label{eqn:op}
\ee
Order parameter is plotted in Fig.~\ref{S3}(b). $ \langle q_6 \rangle$ takes nonzero values after $ \mu \approx 4.26. $  In the disordered phase, all $26$ sublattice densities are on an average  equal and hence $\langle q_6 \rangle=0$. In the sublattice phase one of $Q_A$ or $Q_B$ becomes non-zero, and hence $\langle q_6 \rangle \neq 0$. $\langle q_6 \rangle $  is close to 1 after $\mu \approx 4.26$ implies sublattice-ordered phase.

Since $q_6$ is a scalar, associated Binder cumulant  definition can be written as, 
\begin{equation}
	u_6 = 1- \frac{\langle q_6^4 \rangle}{3\langle q_6^2 \rangle^2} .
\end{equation} The diverging negative peaks in Binder cumulant validates the first order nature of transition. As a further evidence for first order nature of transition, we have plotted pressure in Fig.~\ref{S3}(d). The  pressure loops in $\widetilde{P}$ and  flat curves in $P$ are characteristics of first order transition.

\section*{$7$-NN model}
\begin{figure*}
	\includegraphics[width=\textwidth]{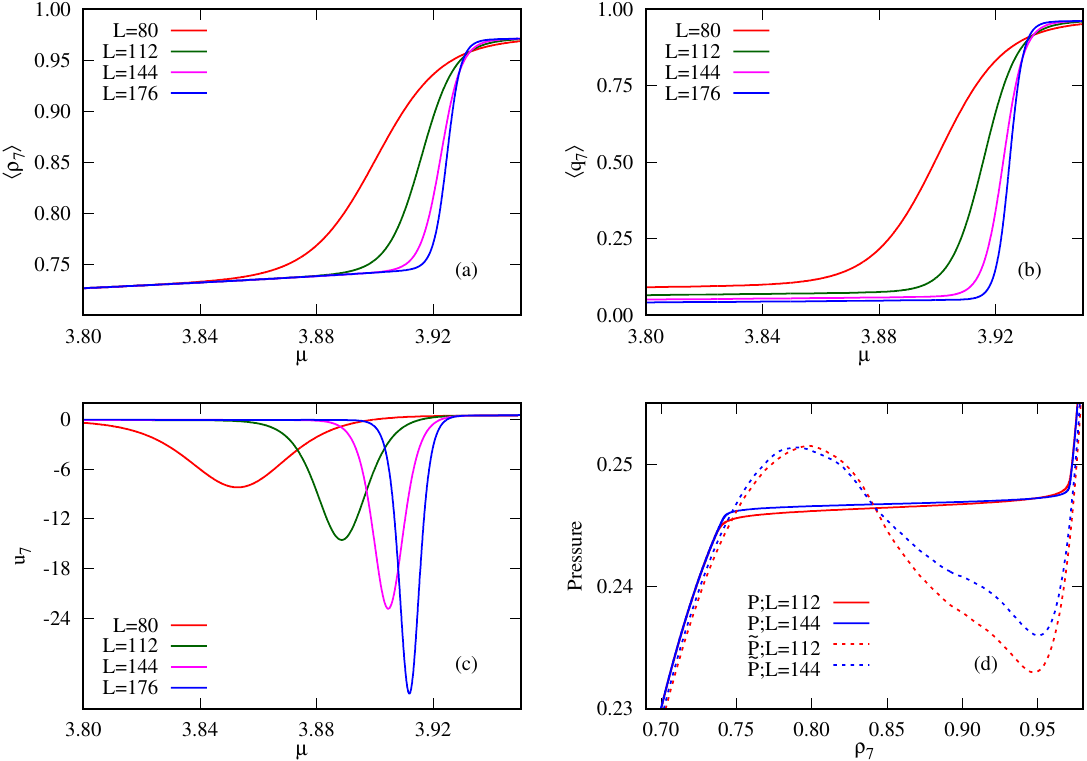}
	\caption{\label{S4} The variation of (a) average density $\langle \rho_7 \rangle$,   (b) average order parameter $\langle q_7 \rangle$ and (c) Binder cumulant $u_7 $  as a function of chemical potential $\mu$ for $7$-NN model. (d) The variation of pressure $ P $ calculated from Eq.~(9) of paper and $\widetilde{P}$ calculated from  Eq.~(11) of paper.  
	}
\end{figure*}

In Fig.~\ref{S4}(a), we plotted average density $\langle \rho_7 \rangle$ against $\mu$. The density rises sharply around $\mu \approx 3.93 $ with the discontinuity becoming sharper with increasing system size, consistent with a first order transition. To show that transition is between disordered phase and sublattice-ordered phase, we have plotted order parameter in Fig.~\ref{S4}(b). In the disordered phase, all $16$ sublattice densities are on an average  equal and hence $\langle q_7 \rangle=0$ as seen in Fig.~\ref{S4}(b). In the sublattice phase one of the sublattice is predominantly occupied and  $\langle q_7 \rangle $  is close to 1 after  $\mu \approx 3.93$ implies sublattice-ordered phase.

The diverging negative peaks in Binder cumulant validates the first order nature of transition. As a further evidence for first order nature of transition, we have plotted pressure in Fig.~\ref{S4}(d). The  pressure loops in $\widetilde{P}$ and  flat curves in $P$ are characteristics of first order transition.